\documentclass{IEEEtran}
\usepackage[utf8]{inputenc}
\usepackage{algorithmic}
\usepackage{graphicx}
\usepackage{textcomp}

\usepackage{color}
\usepackage{placeins}
\usepackage{float}
\usepackage{tabularx,colortbl}
\usepackage{amsmath}
\usepackage{times}
\usepackage{cite}
\usepackage{multirow}
\usepackage{placeins}
\usepackage[RPvoltages,american]{circuitikz}
\usepackage{tikz}
\usetikzlibrary{arrows}
\usetikzlibrary{shapes}
\usetikzlibrary{plotmarks}
\newcommand{\T}[1]{\mathrm{#1}}
\usepackage{pgf}

\usepackage{pgfplots}
\pgfplotsset{compat=newest}
\usetikzlibrary{automata,positioning,arrows}

\begin{document}
\title{DC-Assisted Stabilization of Internal Oscillations for Improved Symbol Transitions in a Direct Antenna Modulation Transmitter}
\author{Danyang Huang, \IEEEmembership{Student Member, IEEE}, Kurt Schab, \IEEEmembership{Member, IEEE}, Joseph Dusenbury, \IEEEmembership{Student Member, IEEE}, Brandon Sluss, \IEEEmembership{Student Member, IEEE}, and Jacob Adams, \IEEEmembership{Senior Member, IEEE}

\thanks{This work was supported by DARPA under grant D16AP00033 and the Department of Navy award N00014-19-1-2713 issued by the Office of Naval Research. Approved for public release; distribution is unlimited.}
\thanks{Danyang Huang, Joseph Dusenbury, Brandon Sluss, and Jacob Adams are with the Department of Electrical and Computer Engineering, North Carolina State University, Raleigh, 
NC 27695 USA (e-mail: dhuang3@ncsu.edu).}
\thanks{Kurt Schab is with the Department of Electrical
and Computer Engineering, Santa Clara University, Santa Clara, CA 95053 USA.}
}

\maketitle

\begin{abstract}
Internal oscillations in switched antenna transmitters cause undesirable fluctuations of the stored energy in the system, reducing the effectiveness of time-varying broadbanding methods, such as energy-synchronous direct antenna modulation. To mitigate these parasitic oscillations, a modified direct antenna modulation system with an auxiliary DC source is introduced to stabilize energy storage on the antenna. A detailed circuit model for a direct antenna modulation system is used to identify the origin of the oscillations and to justify the selection of the DC source. Measured phase shift keyed waveforms transmitted by the modified system show significant increases in signal fidelity, including a 10-20 dB reduction in error vector magnitude compared to a time-invariant system.  Comparison to an equivalent, scalable time-invariant antenna suggests that the switched transmitter behaves as though it has 2-3 times lower radiation Q-factor and 20\% higher radiation efficiency.

\end{abstract}


\section{Introduction}
\label{sec:introduction}

\IEEEPARstart{E}{lectrically} small antennas (ESAs) are highly restricted by a well known lower bound on radiation Q-factor \cite{Chu1948}.  For narrowband and linear time invariant (LTI) systems, radiation Q-factors are directly related to a system's fractional impedance bandwidth, $B$ and radiation efficiency, $\eta$, i.e., 
\begin{equation}
\label{eq:QB}
\eta Q_\mathrm{rad} \sim B^{-1}.
\end{equation}
This suggests that the bandwidth-efficiency product of an LTI antenna system is fundamentally limited by its electrical size. However, these relations only apply to LTI systems, and using a non-LTI network invalidates the relationship between Q-factor and bandwidth, potentially enabling transmission of broadband signals through narrowband antennas~\cite{manteghi2019fundamental,SchabHuangAdams2019_DAMOOK, SchabHuangAdams2020_DAMPSK}. This is particularly desirable at HF and VHF, where the wavelength is long and many antennas of practical physical size are electrically small.

A method generally known as energy-synchronous direct antenna modulation (DAM), switches elements in a small antenna's aperture or matching network to store and release energy on the antenna at specific times.  Switching synchronization and control schemes are selected to enforce specific initial voltage and current conditions within the system, and this energy synchronicity enables extremely rapid transitions between symbol states \cite{Galejs1963}.  Early work on this topic can be traced to very low frequency (VLF) frequency shift keying (FSK) applications \cite{Wolff1957, Johannessen1963, Hartley1971,Vallese1972,Gamble1973} and was later extended to other modulation techniques and higher frequencies \cite{Xu2006, Xu2007, Xu2010,Zhu2014,Manteghi2016,SchabHuangAdams2019_DAMOOK,SchabHuangAdams2020_DAMPSK}.

Recently, the performance of DAM methods tailored to on off keying (OOK) \cite{SchabHuangAdams2019_DAMOOK}, binary phase shift keying (BPSK), and quadrature phase shift keying (QPSK) \cite{SchabHuangAdams2020_DAMPSK} were compared alongside narrowband conventional transmitters. These measurements show dramatic performance improvements compared to an equivalent passive antenna; however, in certain cases, the realized DAM performance deviates significantly from the predictions of simple modeling techniques.  Close study of measurement results indicates that, under certain conditions, undesired oscillations appear within the DAM transmitter circuit during critical time windows when a constant voltage is expected to be held across the antenna terminals \cite{HuangAPS2019, HuangAPS2020}. These undesired oscillations are due to parasitic circuit elements not present in simple models of DAM transmitters and their presence can measurably impede the rapid symbol transitions usually afforded by DAM.

This paper proposes a new DAM transmitter configuration that effectively mitigates the unwanted oscillation in the switched matching network. First, the characteristics of the DAM system are studied using a modified circuit model that includes parasitic elements and predicts the parasitic ringing effects observed in measurement. Secondly, we show that the ringing can be reduced through the introduction of an auxiliary DC voltage source that helps to maintain the desired steady state voltage during symbol transitions. Finally, we conduct over-the-air far field measurements to validate the proposed DC-assisted DAM technique and show that it significantly improves symbol transitions in DAM-QPSK compared to both a conventional LTI transmitter and the original DAM configuration.

\section{Characteristics of DAM Switching Transients}

\label{sec:damcharacter}

Energy-synchronous DAM methods allow for the fast transition between symbol states of differing amplitude (OOK) or phase (PSK) through the use of precisely-timed switched matching networks. In methods based on \cite{Galejs1963}, the DAM transmitter uses an open switch to store voltage on an antenna during the OFF transmission state, and then the switch is closed to resume oscillation and radiation during the ON state. The switching instants are aligned with the peak antenna terminal voltages $v_{\mathrm{a}}(t)$, which maintains the peak steady state voltage on the antenna until the next ON state. The switch is then closed at the instant when the sinusoidal source voltage is also at its peak. In this way, the initial conditions at the switch time exactly match the steady-state conditions and oscillation resumes immediately \cite{Galejs1963,SchabHuangAdams2019_DAMOOK}. 

Ideally, DAM eliminates any transition time between two different states. While detailed experimental studies \cite{SchabHuangAdams2019_DAMOOK,SchabHuangAdams2020_DAMPSK} have shown that DAM reduces symbol transition times and improves signal quality in broadband pulsed OOK and BPSK schemes, little or no benefit was observed in other cases, such as QPSK~\cite{SchabHuangAdams2020_DAMPSK}.  Further experiments indicate that circuit parasitics induce high frequency oscillation during the OFF state, which rapidly alters the energy state on the antenna and impedes the ability to maintain the necessary steady state conditions on the antenna during symbol transitions \cite{HuangAPS2019, HuangAPS2020}. 

\subsection{Transient Analysis of DAM Transmitter}
\label{sec:dam_trans_ana}

\begin{figure}
			\begin{center}
\scalebox{0.65}{
	\begin{circuitikz}[sharp corners,transform shape,line width=0.4pt]
	\ctikzset{inductor=cute, quadpoles style=inline}
	    \node[below] at (-3.5,1) {ON};
		\draw (-2.5,0) to[vsourcesin,l_=$v_{\mathrm{cw}}(t)$] (-2.5,2); 
		\draw (-2.5,2) to[R, l=$R_{\mathrm{s}}$] (-0.5,2);
		\draw (-0.5,2) to[R, l=$R_{\mathrm{sw,ON}}$] (1,2);
		\draw (1,2) -- (1.8,2) to [L, l=$L_{\mathrm{m}}$] (3.8,2) --(5,2) to[open,o-o] (5,0);
		\draw (4.6,2) node[above]{$v_{\mathrm{a}}(t)$} -- ++(0.9,0); 
		\draw (1.8,2) to[capacitor,l=$C_{\mathrm{L1}}$] ++(0,-2);
		\draw (3.8,2) to[capacitor,l=$C_{\mathrm{L2}}$] ++(0,-2);
		\draw  (5.5,2) to[capacitor](7.8,2) coordinate (b);
		\node[right = 4pt] at (6.7,2.3) {$C$};
		\draw (5.5,2) to[capacitor,l=$C_{\mathrm{s}}$] ++(0,-2);
		\draw (b) to[inductor,l=$L$] ++(0,-2);
		\draw (b) -- ++(1,0) to[resistor,l=$R$,*-] ++(0,-2) coordinate (d);
		\draw (8.8,2) node[above] {$v_{\mathrm{rad}}(t)$};
		\node[above] at (5.75,2.4) {$+$};
	    \node[above] at (6.6,2.4) {$v_{\mathrm{C}}(t)$};
		\node[above] at (7.5,2.4) {$-$};
		\draw (d)--(-2.5,0);
		\draw (-2.5,0) node[ground]{};
		\draw (5.5,2);
		
		\draw[dashed,black] (1.4,3) -- (1.4,-0.8);
		\node[below] at (3.2,-0.2) {Matching inductor};
		\node[below] at (-1.5,-0.2) {Source};
		\node[below] at (0.45,-0.2) {SPST};
		\node[below] at (0.45,-0.5) {Switch};
		\draw[dashed,black] (5.1,3) -- (5.1,-0.8);
		\draw[dashed,black] (-0.6,3) -- (-0.6,-0.8);
		\node[below] at (6.8,-0.2) {Antenna};
		
		\node[below] at (3.2,-1.2) {\Large(a)};
\end{circuitikz}}\\
\vspace{0.15in}
\scalebox{0.65}{
	\begin{circuitikz}[sharp corners,transform shape,line width=0.4pt]
	\ctikzset{inductor=cute, quadpoles style=inline}
	    \node[below] at (-2.5,1) {OFF};
		\draw (-1,2.5) to[capacitor,l=$C_{\mathrm{sw}}$] ++(1.5,0) coordinate (a);
		\draw (-1,2.5) -- (-1,1) (a) -- (0.5,1);
		\draw (-1,1) to[R, l=$R_{\mathrm{sw,OFF}}$] (0.5,1);
		\draw (0.5,2) -- (1.8,2) to [L, l=$L_{\mathrm{m}}$] (3.8,2) --(5,2) to[open,o-o] (5,0);
		\draw(-1,2) -- (-1.5,2) -- (-1.5,0);
		
		\draw (4.6,2) node[above]{$v_{\mathrm{a}}(t)$} -- ++(0.9,0); 
		\draw (1.8,2) to[capacitor,l=$C_{\mathrm{L1}}$] ++(0,-2);
		\draw (3.8,2) to[capacitor,l=$C_{\mathrm{L2}}$] ++(0,-2);
		\draw  (5.5,2) to[capacitor](7.8,2) coordinate (b);
		\node[right = 4pt] at (6.7,2.3) {$C$};
		\draw (5.5,2) to[capacitor,l=$C_{\mathrm{s}}$] ++(0,-2);
		\draw (b) to[inductor,l=$L$] ++(0,-2);
		\draw (b) -- ++(1,0) to[resistor,l=$R$,*-] ++(0,-2) coordinate (d);
		\draw (8.8,2) node[above] {$v_{\mathrm{rad}}(t)$};
		\node[above] at (5.75,2.4) {$+$};
	    \node[above] at (6.6,2.4) {$v_{\mathrm{C}}(t)$};
		\node[above] at (7.5,2.4) {$-$};

		\draw (d)--(-1.5,0);
		\draw (-1.5,0) node[ground]{};
		\draw (5.5,2);
		
		\draw[dashed,black] (0.9,3) -- (0.9,-0.8);
		\node[below] at (3,-0.2) {Matching inductor};
		\node[below] at (-0.15,-0.2) {SPST};
		\node[below] at (-0.15,-0.5) {Switch};
		\draw[dashed,black] (5.1,3) -- (5.1,-0.8);
		\node[below] at (6.8,-0.2) {Antenna};
		
		\node [below] at (3.2,-1.3) {$R_{\mathrm{sw,ON}} = 5\ \Omega$, $R_{\mathrm{sw,OFF}} = 27\ M\Omega$, $C_{\mathrm{sw}} = 4$ pF, $L_{\mathrm{m}} = 1750$ nH, $C_{\mathrm{L1}} = 2.9$ pF,};
		
		\node [below] at (2.5,-1.8) {$C_{\mathrm{L2}} = 2.9$ pF, $C_{\mathrm{s}} = 2.7$ pF, $C = 9.3$ pF, $L = 600$ nH, $R = 1950\ \Omega$};
		
		\node [below] at (3.7,-2.7) {\Large(b)};
\end{circuitikz}}\\
\vspace{0.15in}
\scalebox{0.65}{
	\begin{circuitikz}[sharp corners, american voltages, european resistors]
	\tikzstyle{line} = [draw, -latex']
		\node at (5.5,3.3) {$Z_{\mathrm{ant.}}$};
		\path [line] (5,3)-- (6,3);
		
		\draw (-1,2.5) to[resistor,l=$1/sC_{\mathrm{sw}}$] ++(1.5,0) coordinate (a);
		\draw (-1,1) to[R, l=$R_{\mathrm{sw,OFF}}$] (0.5,1);
		\draw (a) -- (0.5,1);
		
		\draw (-1,2.5) -- (-1,1);
		\draw (-1,2)-- (-1.5,2) -- (-1.5,-0.6);
		\draw (-1.5,-0.6) node[ground]{};
		
		\draw (0.5,2) -- (1.2,2) to [resistor, l=$sL_{\mathrm{m}}$] (2.7,2) --(5,2) to[open,o-o] (5,-0.6);		
		\draw (4.8,2) node[above]{$V_{\mathrm{a}}$} -- ++(0.7,0); 
		
		\draw (1.2,2) to[resistor,l=$1/sC_{\mathrm{L1}}$] ++(0,-2.6);
		
		\draw (3,2) to[resistor,l=$1/sC_{\mathrm{L2}}$] ++(0,-1.5);
		\draw (3,-0.6) to [american voltage source, l_=$v_{\mathrm{C_{L2}}}(0)/s$] ++(0,1.1);
		
		\draw  (5.5,2) to[resistor,l=$1/sC$](7.4,2);
		\draw  (8.3,2) coordinate (b) to [american voltage source, l_=$v_{\mathrm{C}}(0)/s$] (7.4,2);
		
		\draw (5.5,2) to[resistor,l=$1/sC_{\mathrm{s}}$] ++(0,-1.5);
		\draw (5.5,-0.6) to[american voltage source, l_=$v_{\mathrm{C_{s}}}(0)/s$] ++(0,1.1);
		
		\draw (b) -- ++(0.2,0) to[resistor,l=$sL$] ++(0,-2.6);
		
		\draw (b) -- ++(1.5,0) to[resistor,l=$R$,*-] ++(0,-2.6) coordinate (d);
		\draw (9.5,2) node[above] {$V_{\mathrm{rad}}$};
		
		\draw (d)--(-1.5,-0.6);
		
		\node[below] at (4.8,-1.5) {\Large(c)};
		
\end{circuitikz}}	\\

			\end{center}
	\caption{Detailed circuit model of DAM transmitter in the (a) ON state, and (b) OFF state for the studied DAM transmitter. (c) The Laplace domain circuit model of DAM OFF state showing initial conditions. The source resistor $R_{\mathrm{s}}$ is assumed to be equal to the resistance of the transmitter in ON state, and an open SPST CMOS switch circuit model is considered in the OFF state.}
	\label{fig:dam-off}	
			\end{figure}
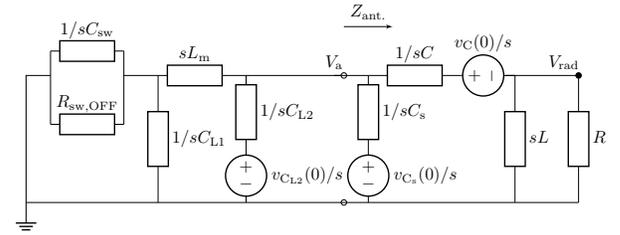

Here we show that a more detailed circuit model of a DAM transmitter including switch and board capacitance, predict the observed ringing behavior. The revised model of the DAM transmitter implemented in \cite{SchabHuangAdams2019_DAMOOK,SchabHuangAdams2020_DAMPSK}, consisting of a series switch, matching inductor, and antenna is shown in Fig.~\ref{fig:dam-off}.  Both ON and OFF state models are shown with their essential parasitic capacitances.    
These parasitics are generated based on measured S-parameters of the circuits used in the DAM transmitters studied in~\cite{SchabHuangAdams2019_DAMOOK,SchabHuangAdams2020_DAMPSK}. Note that we still assume that the switch instantaneously changes between its two states and that any effects due to the transition between states can be neglected. This assumption is valid when the carrier period is much longer than the switching time such that the switch is fast enough to capture the voltage peak, but it may not necessarily apply for higher frequencies or slower switching speeds.  Thus, in this work we select a carrier period that is much longer than the switching time of the switches used in measurement. The voltages $v_{\mathrm{C}}(t)$ and $v_{\mathrm{rad}}(t)$ serve as proxies for the stored energy on the antenna and the fields radiated from the antenna, respectively.

To predict the transient behavior of the DAM circuit, we conduct Laplace analysis to find the antenna terminal voltage $v_{\mathrm{a}}(t)$ in the OFF state, assuming that the circuit is switched from ON to OFF at the instant when the antenna terminal voltage $v_{\mathrm{a}}(t)$ reaches its maximum. Calculation details and complete expressions are given in the Appendix, while only the primary results are presented here.  These analyses yield the following expression for the antenna terminal voltage after switching at $t=0$,
\begin{equation}
\displaystyle v_{\mathrm{a}}(t)\ \approx (V_{\mathrm{ss}}-V_{\mathrm{osc}})+V_{\mathrm{osc}}\mathrm{e}^{-\alpha_{1}t}\cos(\omega_1t),
\label{eq:va_nonideal}
\end{equation}
where
\begin{equation}
\displaystyle V_{\mathrm{ss}}\ \approx\ \frac{1}{2}Q_\mathrm{rad}V_{\mathrm{cw}}
\label{eq:vss_ant}
\end{equation}
is the approximate steady state voltage at the antenna terminal, $Q_\mathrm{rad}$ is the radiation Q-factor of the antenna circuit model, $V_{\mathrm{cw}}$ is the RF source voltage amplitude, $V_{\mathrm{osc}}$, $\omega_1$, and $\alpha_1$ are the amplitude, frequency, and decay constant of the parasitic oscillation at the terminal, respectively.  Specific values of these parameters are computed in the Appendix. This clearly differs from an ideal DAM transmitter as proposed in \cite{Galejs1963}, in which the OFF state terminal voltage should be constant and equal to the peak antenna terminal voltage during steady-state oscillation in the ON state ($V_{\mathrm{ss}}$), i.e.,
\begin{equation}
\displaystyle v_{\mathrm{a, ideal}}(t)\ =\ V_{\mathrm{ss}}.
\label{eq:vss_vaideal}
\end{equation}
Instead, according to \eqref{eq:va_nonideal} the antenna terminal voltage $v_{\mathrm{a}}(t)$  oscillates in a decaying fashion, ultimately converging to a constant but lower voltage. As we will demonstrate in later sections, this alters the performance of the DAM system, depending on when it is switched back into the ON state during the parasitic oscillation cycle.

Following the same procedure, the transient OFF state radiation voltage $v_{\mathrm{rad}}(t)$ of this circuit can be calculated as
\begin{equation}
v_{\mathrm{rad}}(t)\ \approx\ -V_{\mathrm{osc}}^\prime \T{e}^{-\alpha_{1}t}\cos(\omega_1t),
\label{eq:vrad_nonideal}
\end{equation}
where $V_{\mathrm{osc}}^\prime$ is the amplitude of the oscillation at the terminal `$\T{rad}$'. Note that all constants in the expressions \eqref{eq:va_nonideal} and \eqref{eq:vrad_nonideal} are positive, and the negative sign indicates that $v_{\mathrm{rad}}(t)$ is out of phase with the radiation $v_{\mathrm{a}}(t)$ in \eqref{eq:va_nonideal}. In contrast, an ideal DAM transmitter radiates no energy during the OFF state,
\begin{equation}
v_{\mathrm{rad, ideal}}(t)\ =\ 0.
\label{eq:vradideal}
\end{equation}

Thus, considering the parasitics of the circuit, we predict that the DAM antenna will radiate a decaying tone at frequency $\omega_1$, which is not the same as the resonant antenna's tuned frequency and depends on the nominal and parasitic element values. This radiation has the same frequency and decay rate as the oscillation in the antenna terminal voltage $v_{\mathrm{a}}(t)$.


The coefficients in \eqref{eq:va_nonideal} and \eqref{eq:vrad_nonideal} are calculated from the component values in Fig.~\ref{fig:dam-off} and compared with experimental results from a laboratory model of an electrically short monopole antenna at a $28.38$~MHz carrier frequency.
In the experiment, the switch remains closed (ON state) until the system is charged to steady state, then opened when the antenna terminal voltage $v_{\mathrm{a}}(t)$ is at its peak, remaining in the OFF state for $20$ carrier cycles to observe the ringing profile in the OFF state. 

\begin{figure}
\centering
	\includegraphics[width = 1.8in]{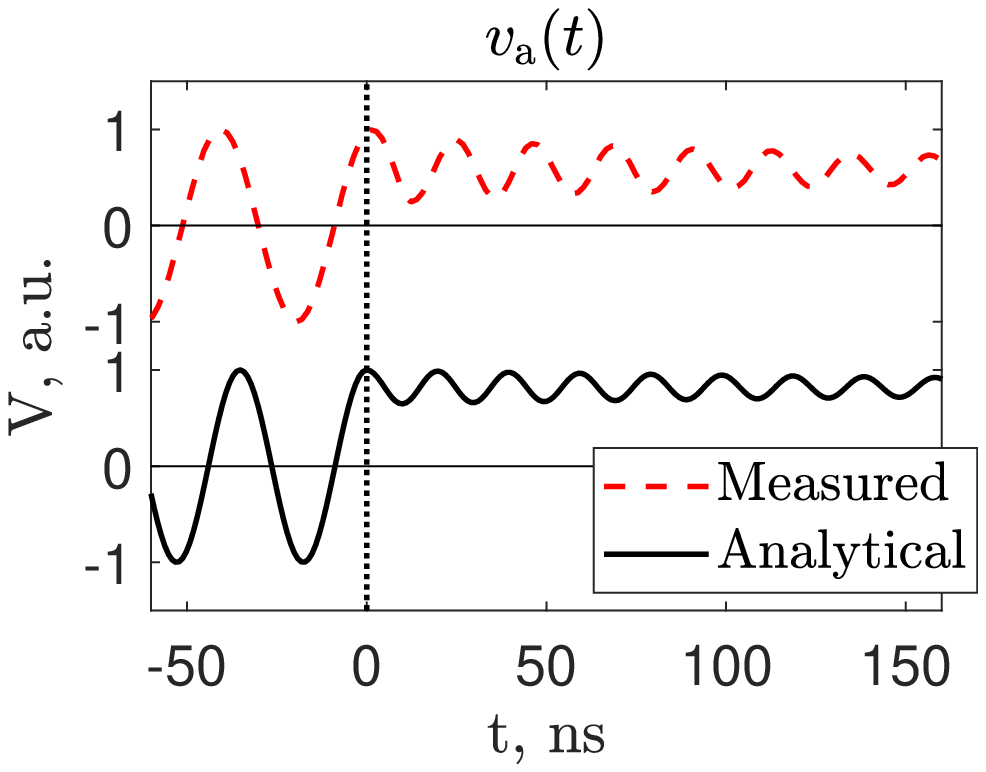}
	\includegraphics[width = 1.62in]{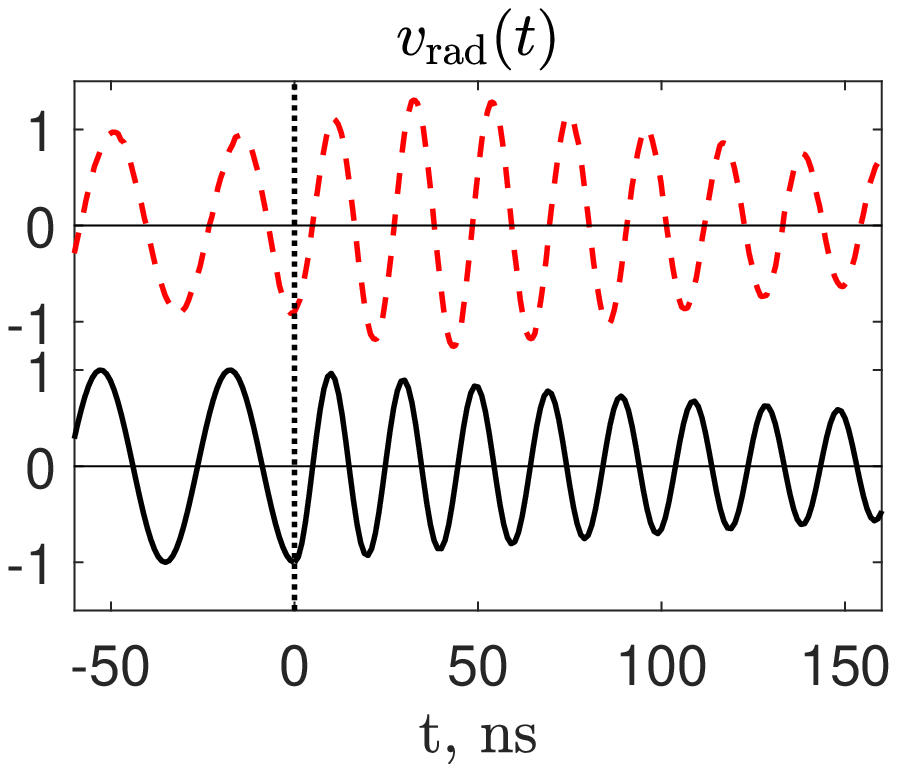}	\\	
		\caption{Terminal voltages of $v_{\mathrm{a}}(t)$ (top) and $v_{\mathrm{rad}}(t)$ (bottom) during the OFF state. A portion of OFF state (nearly $10$ carrier cycles) are shown. The vertical dotted line indicates the ON-OFF transition instant. Results are compared between simulation and the lab test. The carrier frequency of $v_{\mathrm{rad}}(t)$ in both cases are $28.38$~MHz, while $v_{\mathrm{a}}(t)$ in the measurement is de-tuned by the contact-probe to $23.67$~MHz. Results are normalized to their own steady state.}
		\label{fig:dam-off-compare-sim-labtest-derive-va-vrad}
\end{figure}
As can be seen in Fig. \ref{fig:dam-off-compare-sim-labtest-derive-va-vrad}, the measured behavior of the DAM system during the OFF state is well-approximated by the model of Fig.~\ref{fig:dam-off} that includes additional parasitic capacitance. In both $v_{\mathrm{a}}(t)$ and $v_{\mathrm{rad}}(t)$, a large oscillation is observed during the OFF state that decays slowly toward a value below the steady state peak (for $v_{\mathrm{a}}(t)$) or toward zero (for $v_{\mathrm{rad}}(t)$).  Both the frequency $f_\mathrm{rad,OFF}$ and decay rate $\alpha_1$ generally agree between analytic predictions ($f_{\mathrm{rad,OFF}}=50.6$~MHz and $\alpha_1 = 3.6\times10^{6}~\T{s}^{-1}$) and measurement ( $f_{\mathrm{rad,OFF}}=47.1$~MHz and $\alpha_1 = 5.8\times10^{6} ~\T{s}^{-1}$). The circuit model predicts a single, decaying frequency tone with similar characteristics to those observed in measurement. A small deviation between the analytical and measured results is likely caused by the limited complexity of the circuit model used for the antenna, switch, and matching network and slight detuning due to the measurement probe.  


These analytical and measurement results imply that the stored energy in the antenna, represented by $v_{\mathrm{C}}(t) = v_{\mathrm{a}}(t) - v_{\mathrm{rad}}(t)$, is oscillating during the OFF state in a predictable way due to the switch and board capacitances.  These oscillations are critical to the performance of a real DAM transmitter as they change the initial conditions of the circuit at the next OFF-to-ON switching instant and depend highly on the timing of the switching and the specific component parasitics. Moreover, the observed ringing creates undesirable radiation at a secondary frequency which can interfere with other systems. 

\subsection{Ringing effects on DAM QPSK}

\begin{figure}
	\centering
	\includegraphics[trim = 1.7in 3.8in 1.8in 3.9in,clip,width = 2.5in]{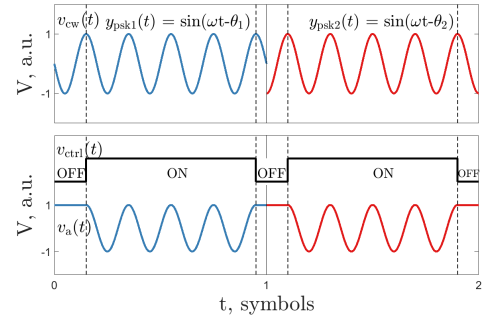}
	\caption{The conceptual DAM PSK signaling scheme, following~\cite{HuangAPS2020,SchabHuangAdams2020_DAMPSK}.  The top panel illustrates the input PSK signal with a $90^\circ$ phase shift.  The bottom panel shows the switch control signal and the antenna terminal voltage, indicating the brief delay period that facilitates rapid, energy-synchronous symbol transitions.
	\label{fig:pskscheme}}
\end{figure}

DAM QPSK is particularly susceptible to performance degradation due to the ringing effects described in the previous section.  In the DAM QPSK scheme studied in \cite{SchabHuangAdams2020_DAMPSK,HuangAPS2020}, the duration of the OFF state depends on the phase transition being made from one symbol to the next, as illustrated in Fig. \ref{fig:pskscheme}. The gap time between the two radiating states is equal to 
$t_\mathrm{{gap}}=\frac{\Delta\theta}{2\pi}T_{\T{c}}$
where $\Delta\theta$ is the phase change (in radians) between states and $T_c$ is the carrier period.

In QPSK there are four such possible transitions ($\Delta\theta\ =\ 0$, $90^\circ$, $180^\circ$, $270^\circ$) and four associated gap delays ($0$, $0.25 T_{\T{c}}$, $0.5T_{\T{c}}$, $0.75T_{\T{c}}$).  Because the ringing frequency is roughly double the carrier frequency, the stored voltage $v_\mathrm{C}(t)$ will have minima near odd multiples of quarter carrier cycles after the switching instant.  Thus, two of the four phase transitions ($90^\circ$, $270^\circ$) appear near the stored energy minima of the parasitic oscillation, see also~\cite{HuangAPS2019, HuangAPS2020}. 

\begin{figure}			
	\begin{center}
		\includegraphics[trim = 1.4in 4.2in 1.8in 4in,clip,width = 1.8in]{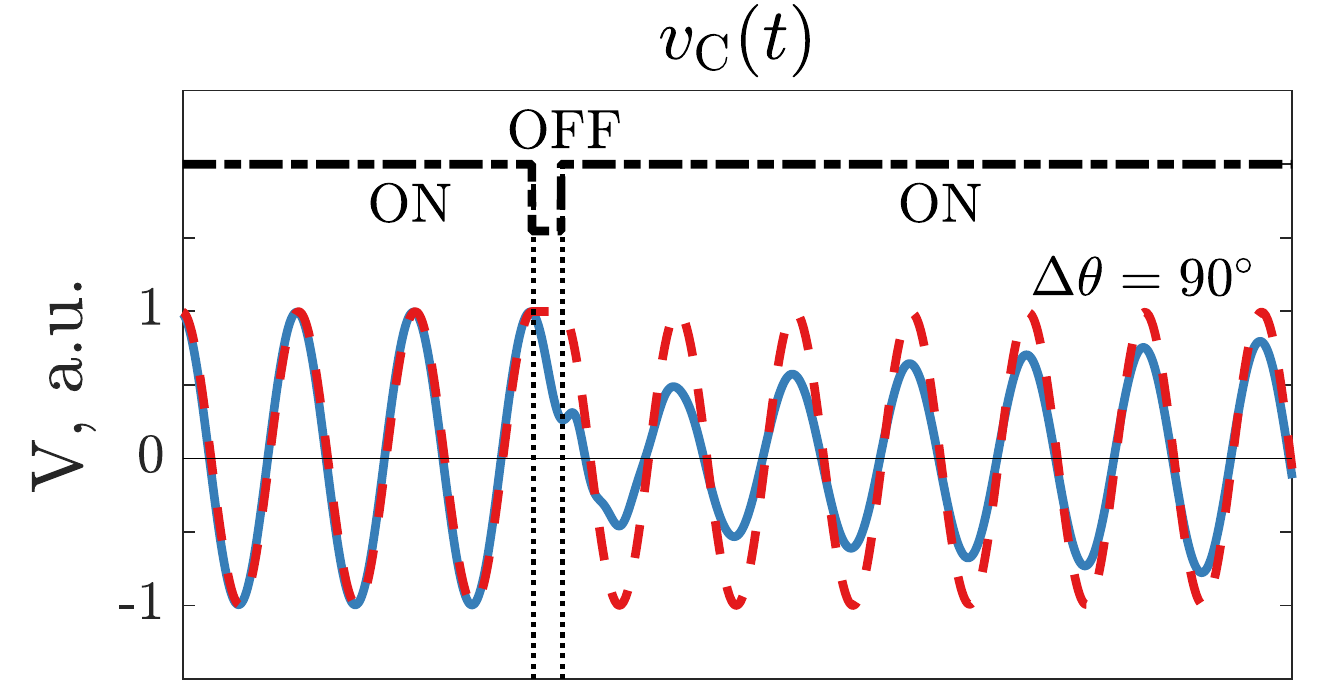}
		\includegraphics[trim = 1.8in 4.2in 1.9in 4in,clip,width = 1.635in]{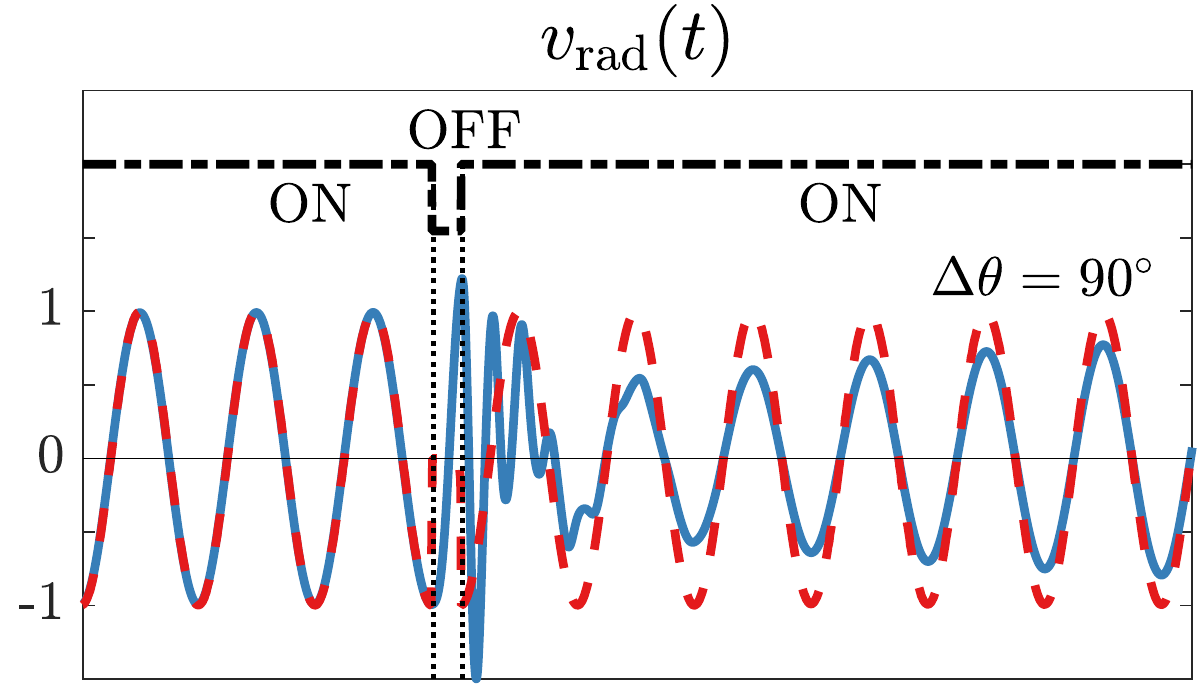}\\
		\includegraphics[trim = 1.4in 3.9in 1.8in 4.15in,clip,width = 1.8in]{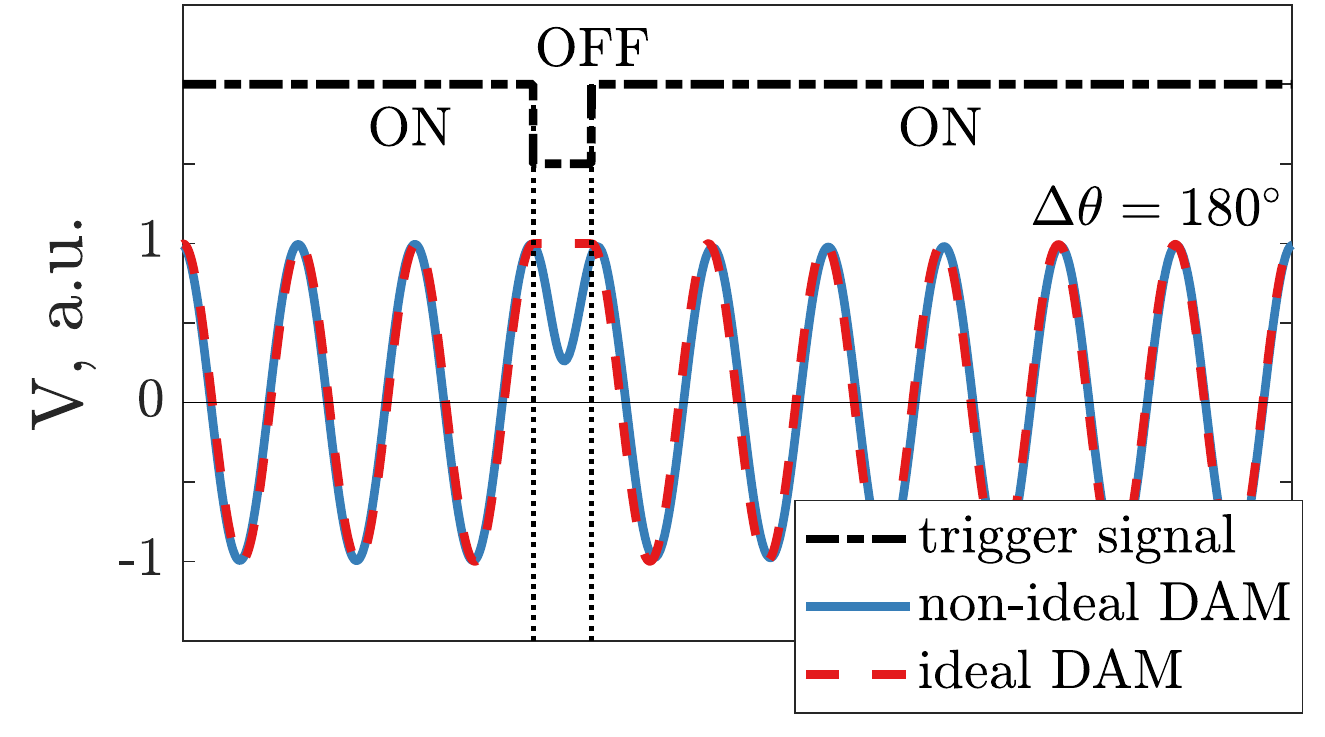}
		\includegraphics[trim = 1.8in 3.9in 1.9in 4.15in,clip,width = 1.635in]{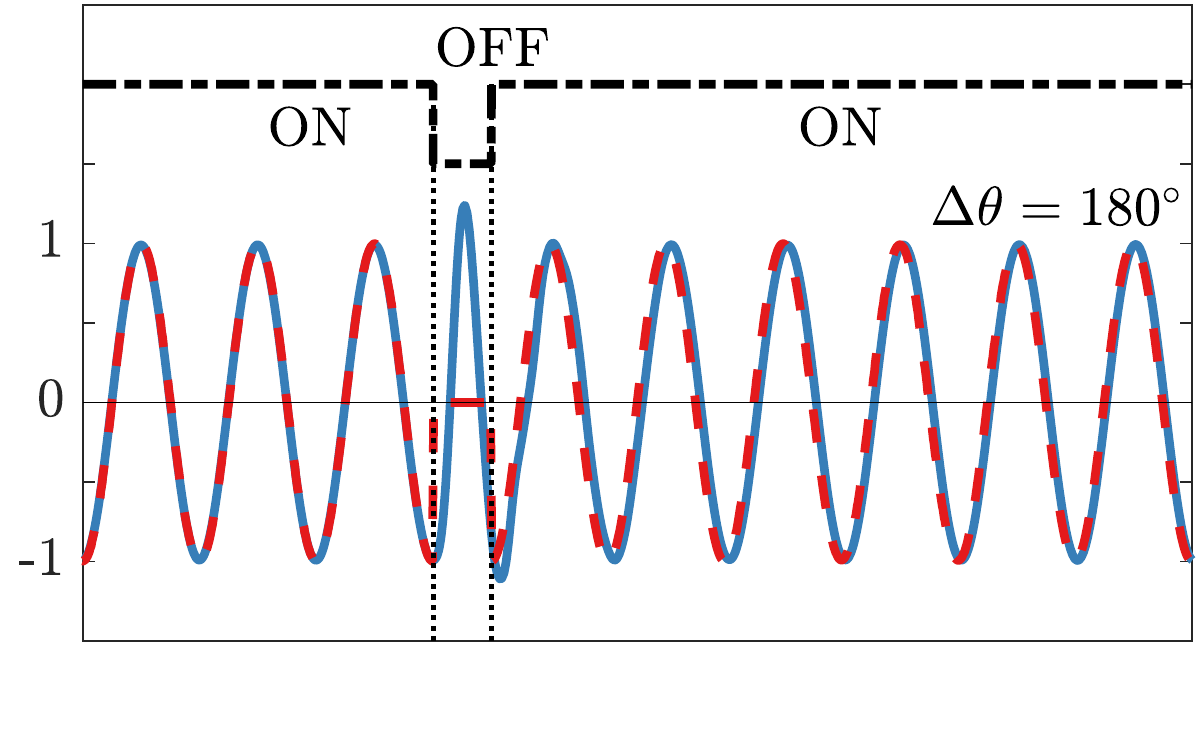}\\
		\includegraphics[trim = 1.4in 3.9in 1.8in 4.15in,clip,width = 1.8in]{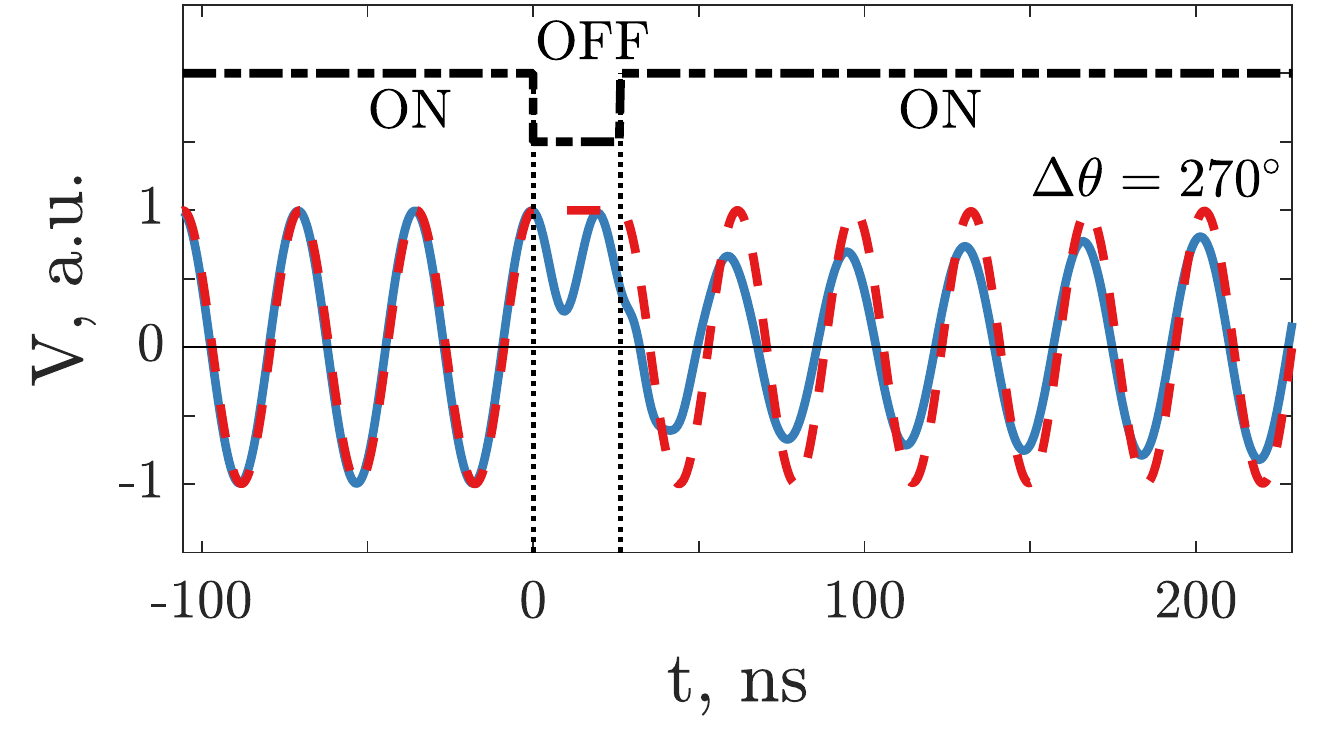}
		\includegraphics[trim = 1.8in 3.9in 1.9in 4.15in,clip,width = 1.635in]{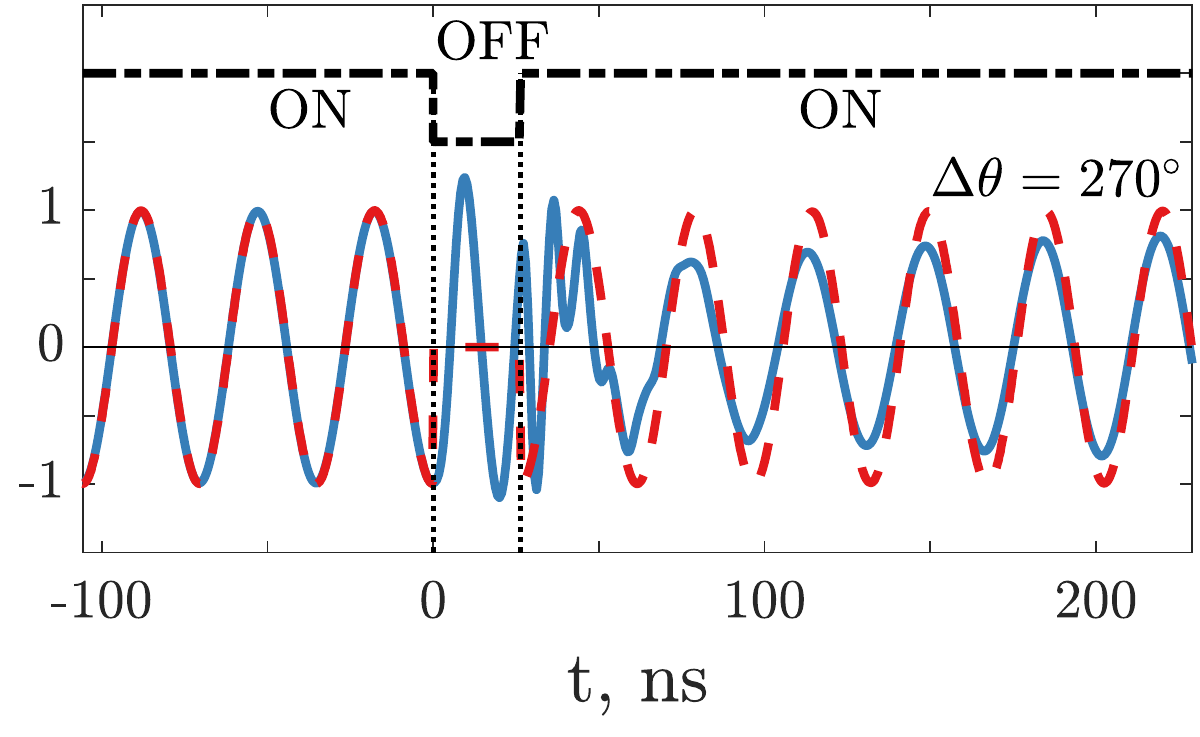}\\
	\caption{Normalized time-domain signal $v_{\mathrm{C}}(t)$ (left column) and $v_{\mathrm{rad}}(t)$ (right column) when a sequence of ``ON-OFF-ON" is transmitted. The switch-off time is (top) $0.25T_{\T{c}}$, (middle) $0.5T_{\T{c}}$, and (bottom) $0.75T_{\T{c}}$, which correspond to $90^\circ$, $180^\circ$, and $270^\circ$ phase shifts. Dashed lines show the ideal DAM signal of each case.}
	\label{fig:dam-off-SPST-timedomain-signal_va-vrad}
	\end{center}
\end{figure}

To observe the effect of switching during these instants of low stored voltage, simulations of a single DAM-QPSK phase transition are carried out using a transient circuit solver (Advanced Design System). First, the system is driven in the ON state with the switch closed and allowed to charge to the steady state.  We then briefly open the switch (OFF) when $v_{\mathrm{a}}(t)$ is at maximum and close it again after a gap delay corresponding to a chosen phase transition.
The stored voltage $v_{\mathrm{C}}(t)$ and radiated field $v_{\mathrm{rad}}(t)$ are plotted in Fig.\ \ref{fig:dam-off-SPST-timedomain-signal_va-vrad} for three different phase transitions.

The simulations in Fig.~\ref{fig:dam-off-SPST-timedomain-signal_va-vrad} illustrate an important problem that arises due to the oscillation of the OFF state voltage of the DAM transmitter. Because the $90^\circ$ and $270^\circ$ phase transitions require switching at instants that correspond with minima of the OFF state voltage oscillation, the low initial voltage at these transitions results in a slow charging time. Relative to the ideal DAM case, these two transitions show a slow charging envelope, while the $180^\circ$ transition, with a gap delay that aligns with a ringing peak, is very closely aligned with the ideal. Thus, in these QPSK cases, the circuit parasitics and OFF state ringing significantly reduce the DAM method's advantages over a conventional LTI transmitter. 


\section{DAM with an Auxiliary DC Source}

\begin{figure}
	\centering
	\scalebox{0.6}{
		\begin{circuitikz}[scale=1,transform shape,line width=0.65pt]
	\ctikzset{inductor=cute, quadpoles style=inline}
		\draw (0,-1.5) node[ground]{} ++(0,1);
		\draw (0,-1.5) to[vsourcesin,l_=$v_{\mathrm{RF}}$] ++(0,1.2-0.01) (0, -0.315) -- (1,-0.315) (1.5,0) node[spdt,rotate=180]{}  (2,0) to [inductor,l^=$L_{\mathrm{m}}$] ++(1.5,0)  (3.5,0) node[txantenna]{} (6,1) node[below]{$E_{\mathrm{ff}}$};
		\draw (4.9,-1.47) node[ground]{} ++(0,1);
		\draw (4.9,-0.1) to[open,-o, v=$V_{\mathrm{a}}$] (4.9,-1.4);
		\draw (-1.5,-1) node[ground]{} ++(0,1);
		\draw (-1.5,-1) to[american voltage source,l^=$V_{\mathrm{DC}}$] (-1.5, 0.32);
		\draw (-1.5, 0.31) -- (0.95, 0.31);
	\end{circuitikz}}	\\
	
	\scalebox{0.65}{
	\begin{circuitikz}[sharp corners, american voltages, european resistors]
	\tikzstyle{line} = [draw, -latex']
		\node at (5.5,3.3) {$Z_{\mathrm{ant.}}$};
		\path [line] (5,3)-- (6,3);

		\draw (-0.8+0.7+1,2) to[american voltage source, l^=$V_{\mathrm{DC}}/s$] ++(1.7,0) coordinate (a);
		
		
		\draw (-0.8+0.7+1, 2) -- (-0.8+0.7+1, -0.6);
		\draw (-0.8+0.7+1,-0.6) node[ground]{};
		
		\draw (1.2+0.4+1,2) to [resistor, l=$sL_{\mathrm{m}}$] (2.7+0.7+1,2) --(5,2) to[open,o-o] (5,-0.6);		
		\draw (4.8,2) node[above]{$V_{\mathrm{a}}$} -- ++(0.7,0); 
		
		
		
		\draw  (5.5,2) to[resistor,l=$1/sC$](7.4,2);
		\draw  (8.3,2) coordinate (b) to [american voltage source, l_=$v_{\mathrm{C}}(0)/s$] (7.4,2);
		
		
		\draw (b) -- ++(0.2,0) to[resistor,l=$sL$] ++(0,-2.6);
		
		\draw (b) -- ++(1.5,0) to[resistor,l=$R$,*-] ++(0,-2.6) coordinate (d);
		\draw (9.5,2) node[above] {$V_{\mathrm{rad}}$};
		
		\draw (d)--(-0.8+0.7+1,-0.6);
		
\end{circuitikz}}	\\

	\caption{Simplified schematic of the DAM transmitter with auxiliary DC source (top).  When the switch is in the lower (ON) position, the RF source is connected to the antenna and the signal is radiating.  When the switch is in the upper position (OFF), the auxiliary DC source is connected to the antenna to maintain the required voltage at $V_\mathrm{a}$ and radiation is paused. (bottom) Simplified Laplace domain model of DC-DAM OFF state showing initial conditions.}
	\label{fig:scheme-dam-dc}	
\end{figure}
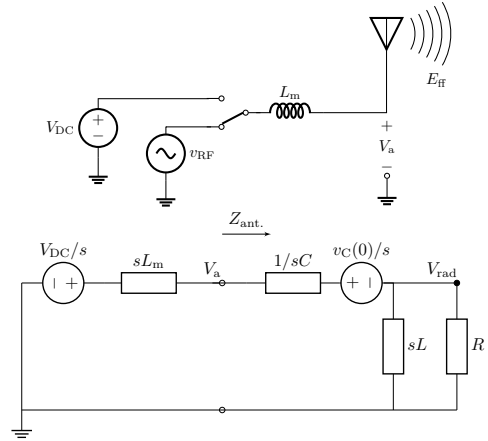

As discussed in the previous section, unwanted oscillations are generated during the DAM transmitter's OFF state due to parasitics within the transmitter and switch.  This leads to spurious radiation and an inability to maintain the proper initial conditions required for fast symbol transitions. In this section, we mitigate these effects by introducing an auxiliary DC source that stabilizes the antenna voltage during the OFF state.  This concept, referred to here as \emph{DC-assisted DAM}, is illustrated in Fig. \ref{fig:scheme-dam-dc}.  DC sources have been used in other directly modulated systems but for different purposes than those here (\textit{e.g}, to inject power in place of an RF source~\cite{Salehi2014} and to pre-charge a DAM receiver~\cite{WangTMTT2007}).

In the proposed configuration, a DC voltage is impressed on the antenna during each OFF state to stabilize the antenna terminal voltage. The value of the auxiliary DC voltage is critical. To select it, we consider a modified OFF state circuit model in which the switch and board parasitics to the left of the inductor in Fig.~\ref{fig:dam-off} are shorted out by the low impedance DC source path.  We further simplify the model by neglecting the charge stored in the two parasitic capacitances between the node ``$\mathrm{a}$'' and ground, which is small compared to the charge on the primary antenna capacitor $C$.  Thus, the addition of the DC source can be modeled by the simplified OFF-state circuit in Fig~\ref{fig:scheme-dam-dc}.  From this model, it is evident that selecting
\begin{equation}
\displaystyle V_{\mathrm{DC}} = v_{\mathrm{C}}(0) =  V_{\mathrm{ss}} - v_\mathrm{rad}(0) \approx V_{\mathrm{ss}}
\label{eq:Vdc}
\end{equation}

\noindent will cancel the residual stored energy in $C$ and produce zero current flow, consequently eliminating the parasitic ringing and radiation.  This result arises from several approximations, and in practice small oscillations are observed.  However, these effects are significantly reduced compared to the case when the antenna is left open-circuited in the OFF state, i.e., the ``open-circuit DAM'' (OC-DAM) described in Sec.~\ref{sec:dam_trans_ana}.

\begin{figure}	
    \centering
		\includegraphics[width = 1.8in]{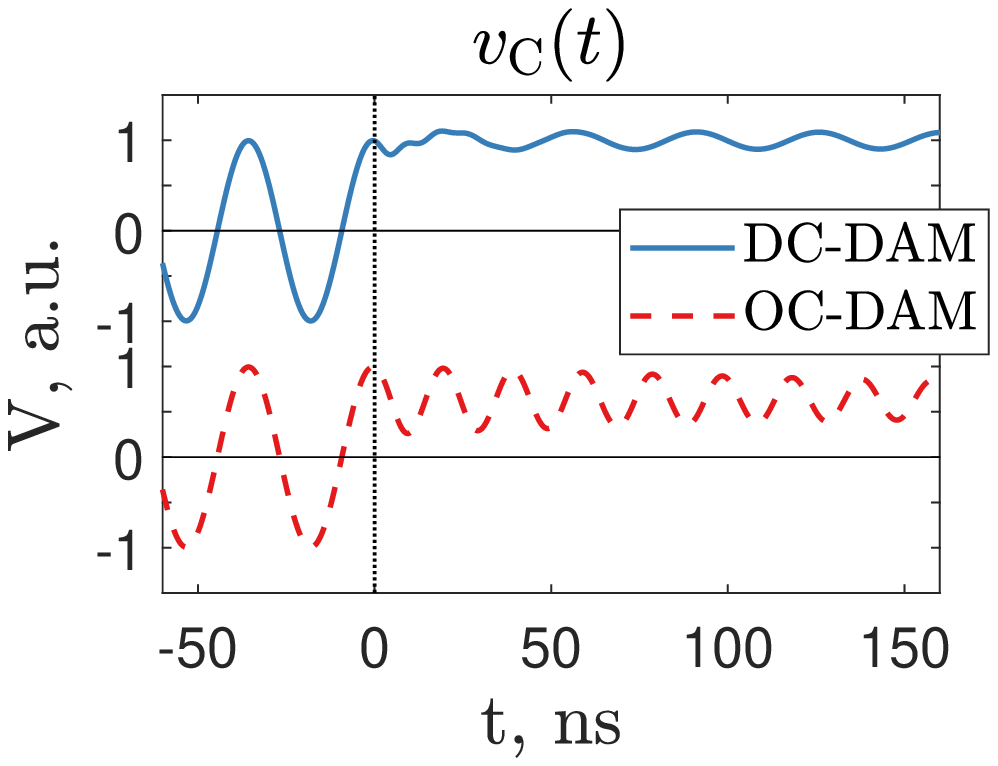}
	\includegraphics[width = 1.62in]{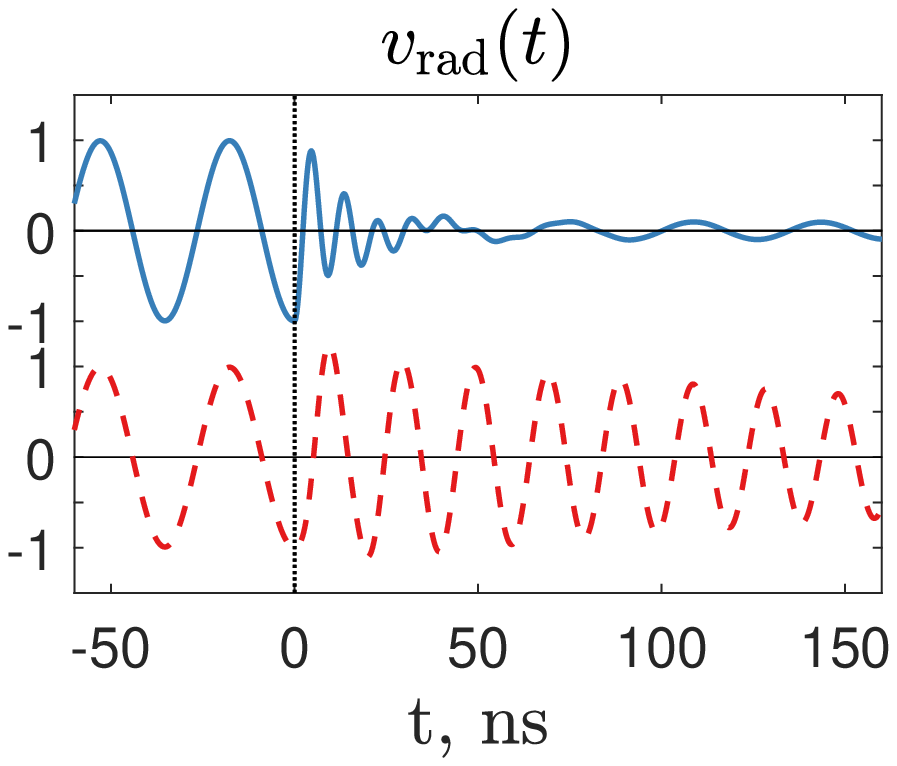}	\\
    	\caption{Normalized simulated $V_{\mathrm{C}}$ (left) and $V_{\mathrm{rad}}$ (right) of DC-assisted DAM (DC-DAM) and open-circuit DAM (OC-DAM) during the OFF state. The dotted line indicates the ON-OFF  transition instant.}
	\label{fig:dc-dam-sim-on-off}
\end{figure}

To demonstrate the effect of the auxiliary DC source, the ``ON - OFF'' sequences described in Section \ref{sec:dam_trans_ana} are simulated using the circuit in Fig.~\ref{fig:scheme-dam-dc} including all switch, inductor, and antenna parasitics. The simulations shown in Fig.~\ref{fig:dc-dam-sim-on-off} comparing the DC-assisted DAM case with the open-circuit DAM scheme.
We observe that the oscillation in both $v_{\mathrm{C}}(t)$ and $v_{\mathrm{rad}}(t)$ are both greatly reduced, while the DC antenna voltage in $v_{\mathrm{C}}(t)$ is held closer to the steady state level during the entire OFF state. While some OFF-state variation remains visible in both quantities, the oscillation frequency of the OFF-state DC-assisted DAM remains at the carrier frequency as the antenna is shorted by the DC source. This in contrast to the higher frequency transients observed in the open-circuit DAM cases.

QPSK transitions with an auxiliary DC source are simulated using the circuit in Fig.~\ref{fig:scheme-dam-dc} including all switch, inductor, and antenna parasitics. As before, three phase transitions are simulated, each with a different gap delay. 
\begin{figure}			
	\begin{center}
		\includegraphics[trim = 1.4in 4.2in 1.8in 4in,clip,width = 1.805in]{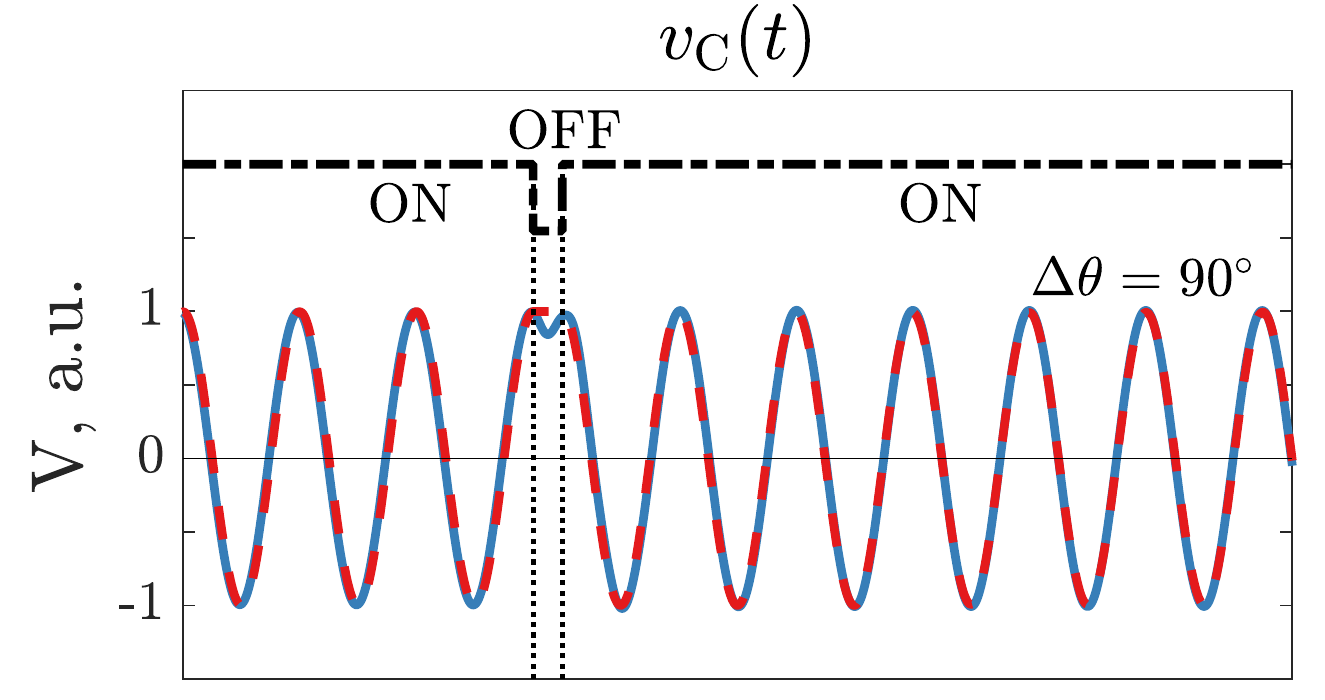}
		\includegraphics[trim = 1.8in 4.2in 1.9in 4in,clip,width = 1.635in]{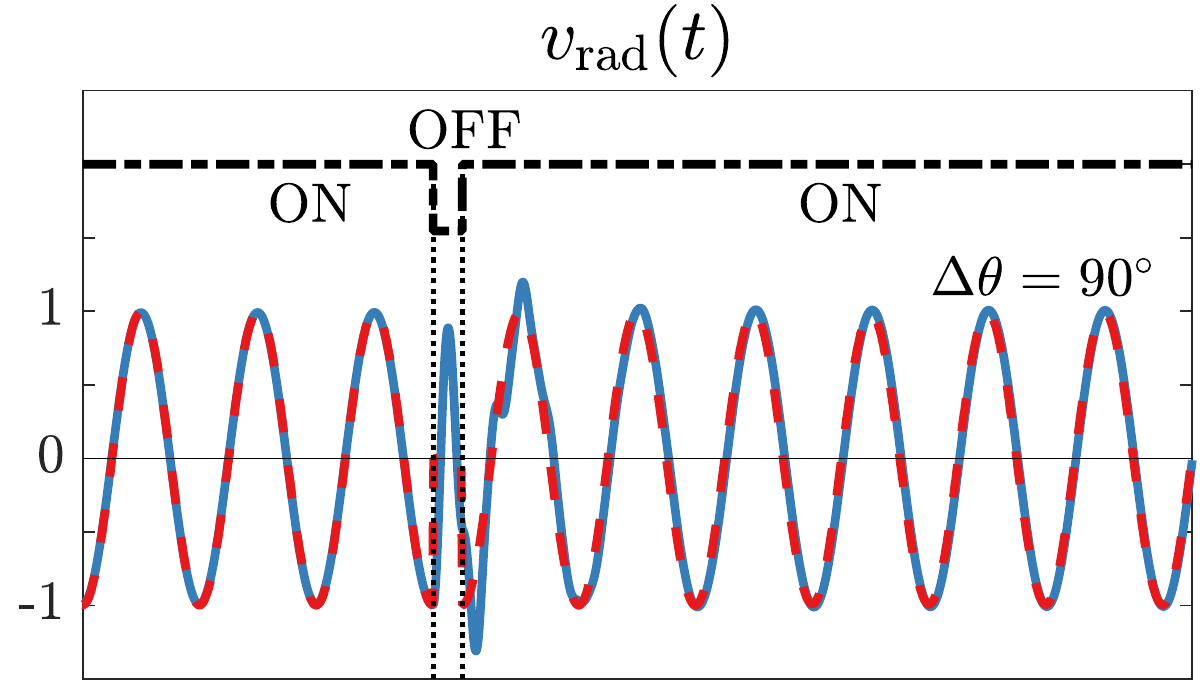}\\
		\includegraphics[trim = 1.4in 3.9in 1.8in 4.15in,clip,width = 1.805in]{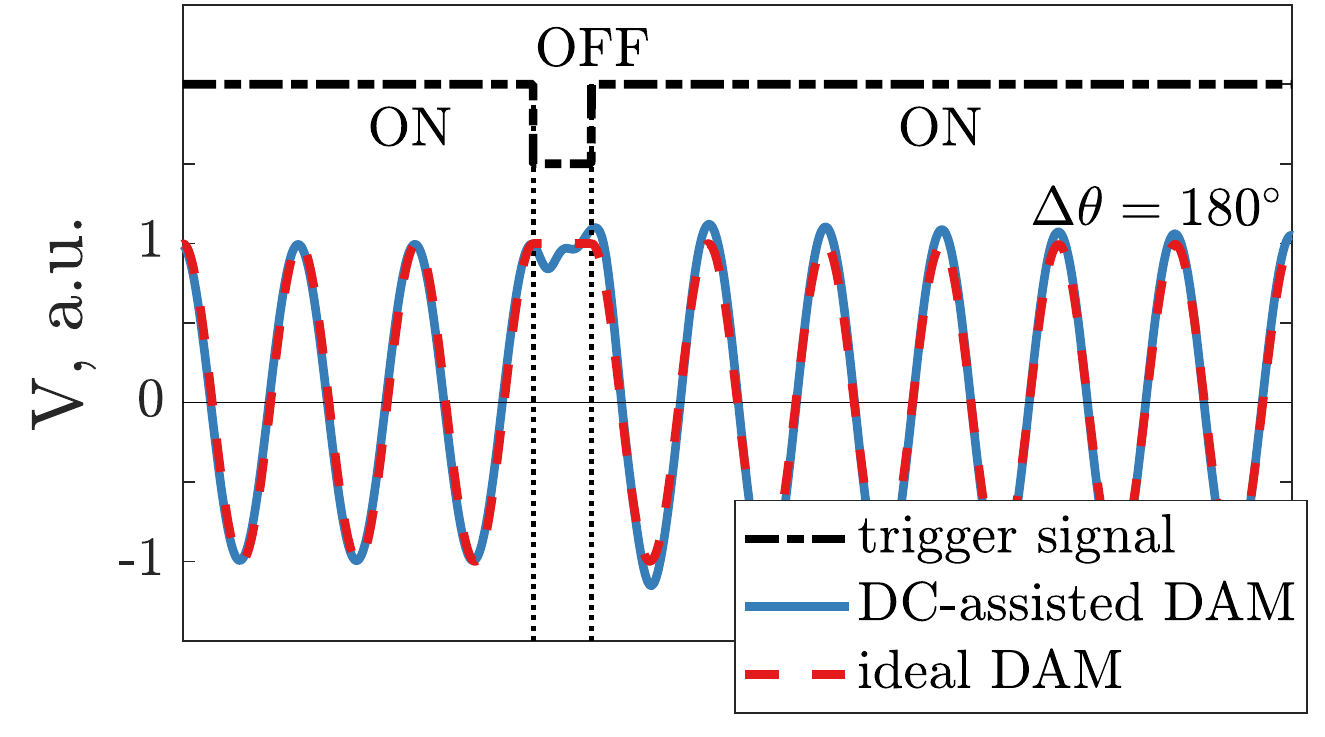}
		\includegraphics[trim = 1.8in 3.9in 1.9in 4.15in,clip,width = 1.635in]{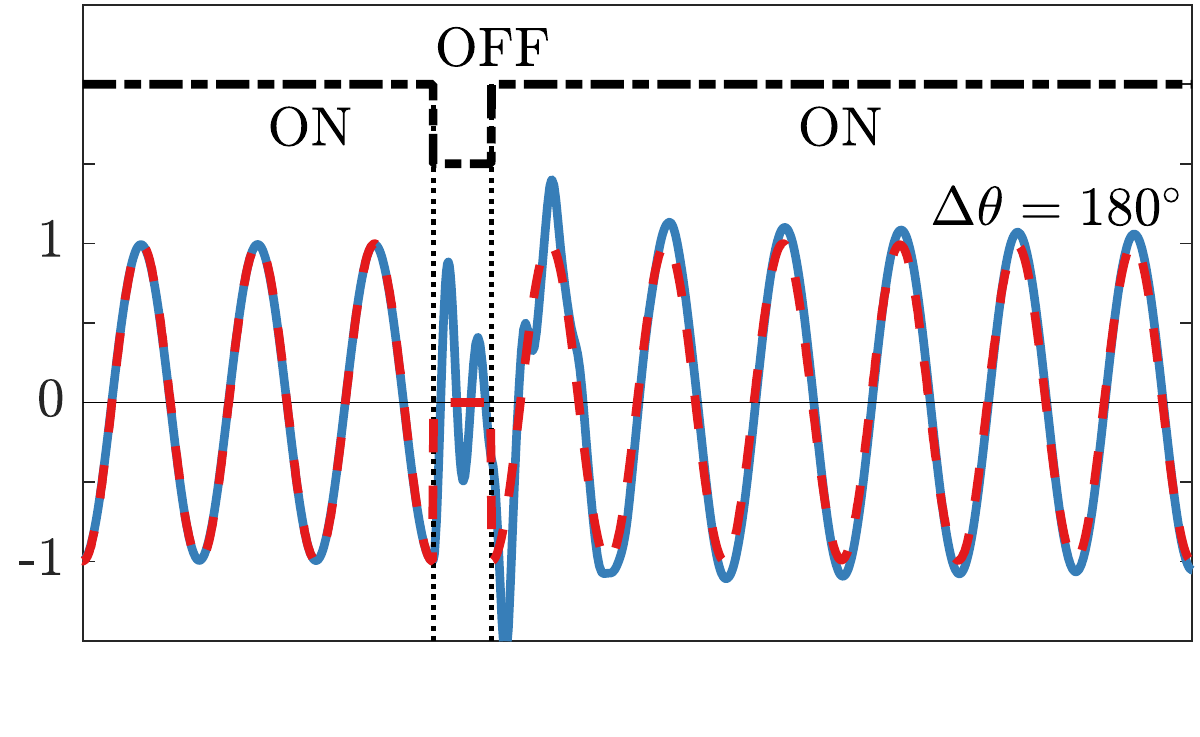}\\
		\includegraphics[trim = 1.4in 3.9in 1.8in 4.15in,clip,width = 1.805in]{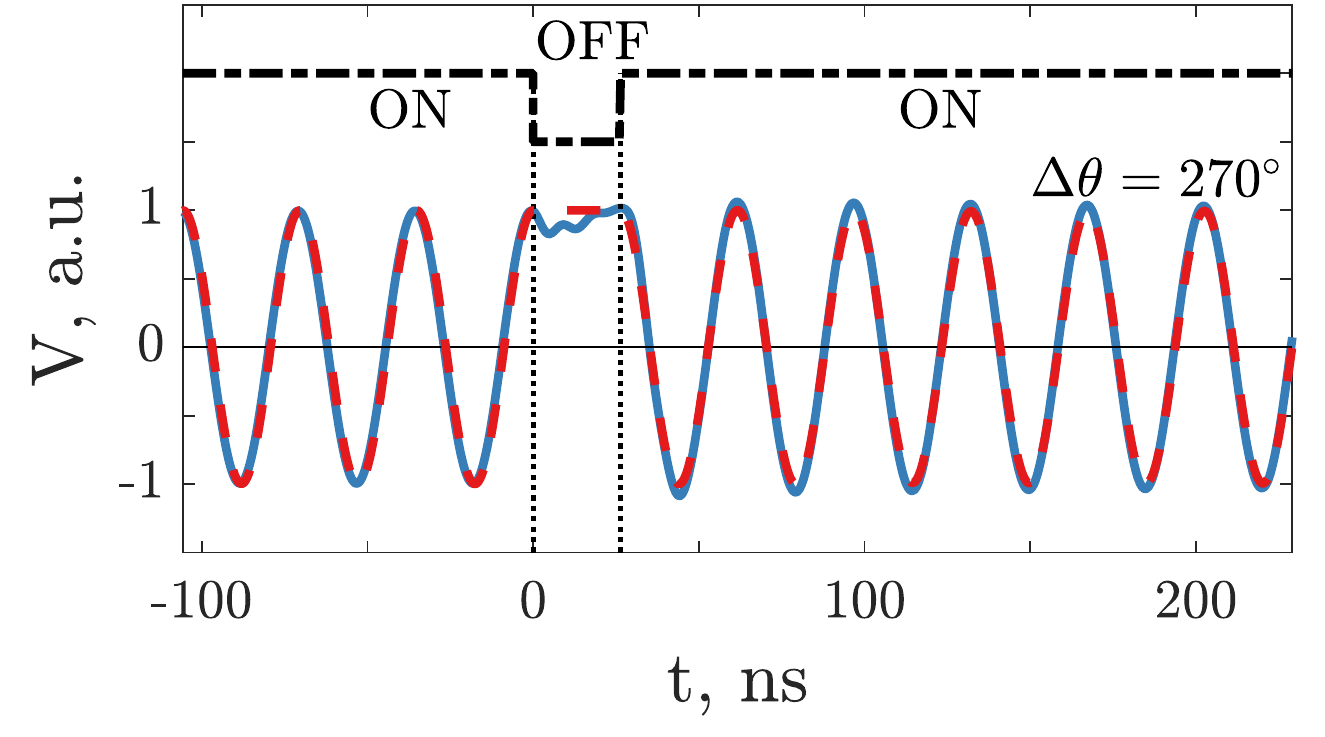}
		\includegraphics[trim = 1.8in 3.9in 1.9in 4.15in,clip,width = 1.635in]{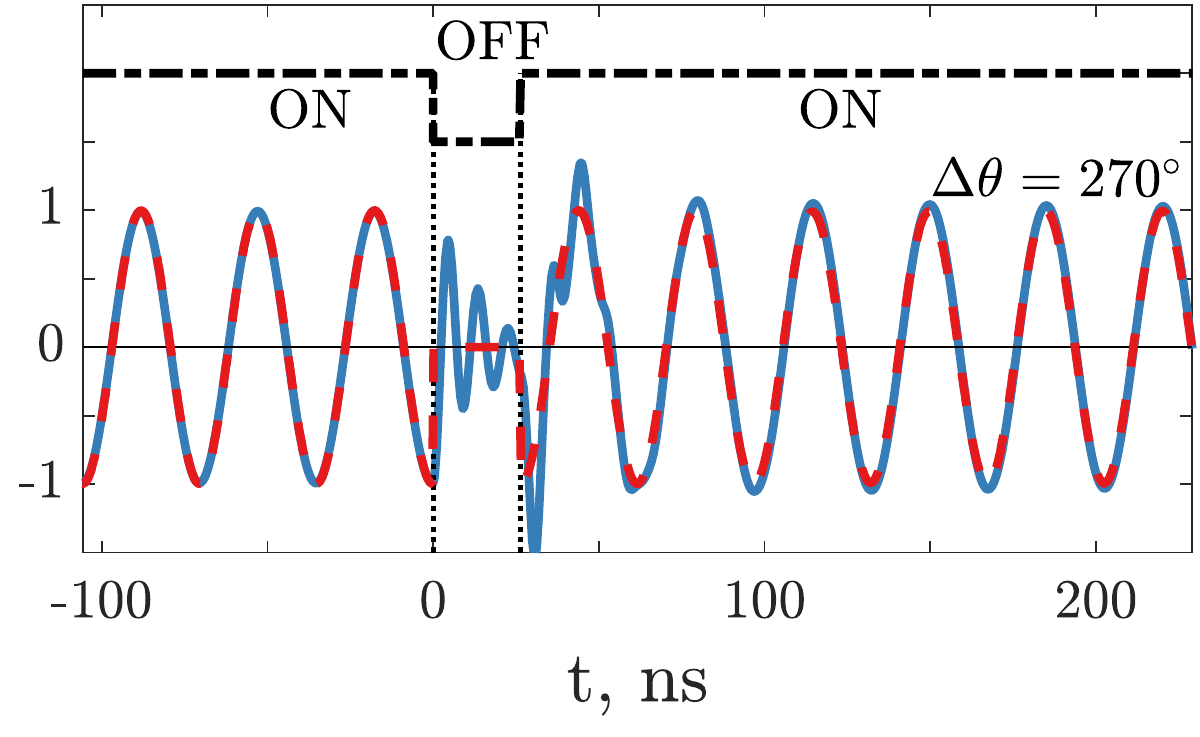}\\
	\caption{Normalized time-domain signal $v_{\mathrm{C}}(t)$ (left column) and $v_{\mathrm{rad}}(t)$ (right column) of DC-assisted DAM when a sequence of ``ON-OFF-ON" is transmitted. The switch-off time is (top) $0.25T_{\T{c}}$, (middle) $0.5T_{\T{c}}$, and (bottom) $0.75T_{\T{c}}$, which correspond to $90^\circ$, $180^\circ$, and $270^\circ$ phase shifts. Dashed lines show the ideal DAM signal of each case.}
	\label{fig:dc-dam-sim-diff-datarate}	
	\end{center}
\end{figure}

The simulations of this DC-assisted DAM case shown in Fig.~\ref{fig:dc-dam-sim-diff-datarate} contrast with the open-circuited DAM case in Fig.~\ref{fig:dam-off-SPST-timedomain-signal_va-vrad}.  We observe that the oscillation in $v_{\mathrm{C}}(t)$ during the OFF state is nearly eliminated by the auxiliary DC source, and the voltage stored on the antenna is maintained at the steady state level for the duration of the OFF delay. Although the radiated field $v_{\mathrm{rad}}(t)$ exhibits transients after state transitions, the carrier returns to its steady state amplitude almost immediately after the switch is closed. As anticipated, the auxiliary DC source cancels the undesired oscillation during the OFF state and provides more stable initial conditions for the antenna at the switching instant, leading greatly reduced transition time between symbols.

\begin{figure}			
	\begin{center}
    \includegraphics[trim = 1.5in 3.3in 1.8in 3.5in,clip,width=3in]{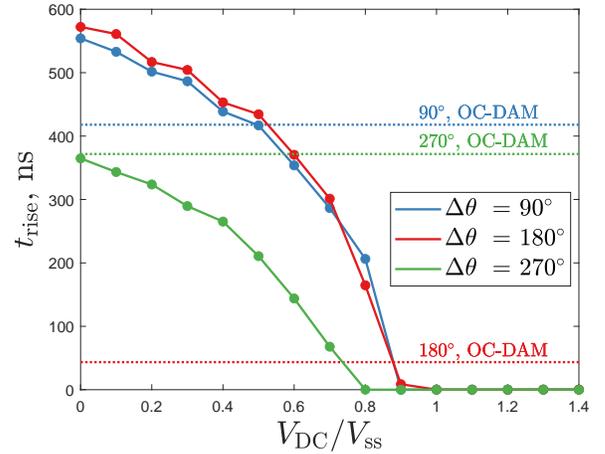}\\
	\caption{Comparison of DC-assisted DAM performance (curves with markers)  against open-circuit DAM (OC-DAM, horizontal lines) described by the rise time to $95\%$ of the steady state of the radiated signal after the switch-off time corresponding to $90^\circ$, $180^\circ$, and $270^\circ$ phase shifts.}
	\label{fig:dc-dam-sim-diff-vdc}	
	\end{center}
\end{figure}

A parametric study of varying DC voltage levels is carried out to determine if the condition in (\ref{eq:Vdc}) is in fact the optimal choice.  The DC signal $V_{\T{DC}}$ is varied in a phase transition experiment in which the switch-off delays are $0.25T_{\T{c}}$, $0.5T_{\T{c}}$, and $0.75T_{\T{c}}$ (representing 90$^\circ$, 180$^\circ$, and 270$^\circ$ phase shifts). The received signal is down-converted and results shown in Fig.~\ref{fig:dc-dam-sim-diff-vdc} plot the rise time to $95\%$ of the steady state amplitude of the radiated signal after the switch-off time. We observe that when $V_{\T{DC}}$ is low, both the rise time of three cases are slow, bringing no benefit to the system. Particularly, if $V_{\T{DC}}\ =\ 0~\mathrm{V}$, the rise time of the signal in each transitions are roughly equivalent to that of a conventional LTI transmitter. When $V_{\T{DC}}$ reaches the steady state of the $v_{\T{C}}(t)$, i.e., $V_{\T{DC}}/V_{\T{ss}} = 1$, all three transients have nearly instantaneously rising envelopes. While increasing the DC signal to be above the steady state value $V_{\T{ss}}$ does not further improve the rise time, the choice of the DC signal to be $V_{\T{DC}}=V_{\T{ss}}$ is considered as the optimal solution. The effect of the DC-source on QPSK signal quality is studied experimentally in the next section.

\section{Experimental Characterization of DC-Assisted DAM}

\subsection{Experimental Setup}
\begin{figure}
    \centering
    \input{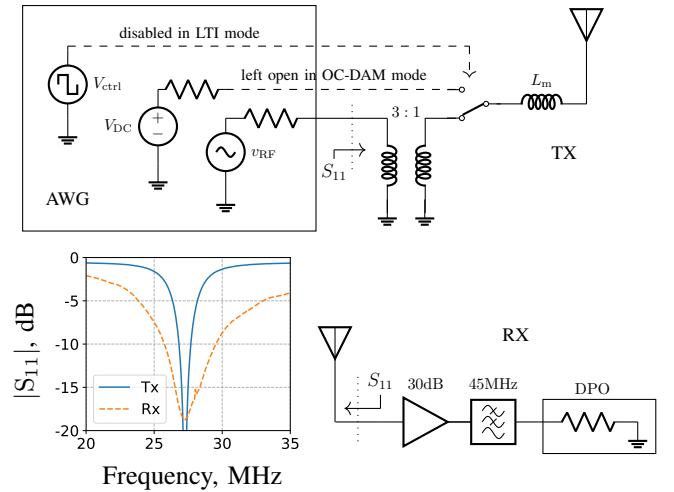}
\caption{Schematic of experimental transmitter (top), receiver (bottom right).  Also shown is the tuned transmitter and receiver reflection coefficients as seen in LTI mode (bottom left). Signals are generated by the arbitrary waveform generator (AWG) and sampled by the oscilloscope (DPO).}
     \label{fig:block-ota}
 \end{figure}

To test the proposed DC-assisted DAM transmitter using QPSK modulation, a transmitter and receiver are set up in an open area on North Carolina State University's campus in an identical manner to \cite{SchabHuangAdams2019_DAMOOK, SchabHuangAdams2020_DAMPSK}. A schematic of the far field measurement system is shown in Fig.~\ref{fig:block-ota}.
 The transmitting antenna is an electrically-short $0.94$~m tall monopole antenna (height of $0.085\lambda$ at carrier frequency in the ISM band at $27.23$ MHz, $Z_\mathrm{in}=17-\mathrm{j}550\ \Omega$) tuned to resonance at $27.23$~MHz using a series variable inductance and matched with a $3:1$ transformer. The receiving quarter-wave monopole antenna is placed approximately $60$~m (equivalently $5.45$ wavelengths at $27.23$~MHz) away from the transmitter and connected to a lowpass filter (KR Electronics, KR2805) and a low-noise amplifier (Minicircuits LNA-530). Both antennas are placed perpendicular to the earth and driven against separate ground planes composed of wire radials laid over the earth ground. The received signal is sampled by a 4 GHz oscilloscope (Tektronix DPO70404C). The antennas are selected so that the transmit antenna's 10~dB return loss bandwidth~($3.2\%$), see Fig.~\ref{fig:block-ota}, is much narrower than the broadband QPSK signal's first-null fractional bandwidth ($40\%$, $66\%$ for QPSK with $N=5,~ 3$, respectively, where $N=T_{\T{s}}/T_{\T{c}}$ and $T_{\T{s}}$ and $T_{\T{c}}$ are the symbol and carrier periods, respectively) as well as the receive antenna's bandwidth ($15.7\%$).  Both antennas are supported by PVC masts and eight radials over earth ground improve the ground plane performance. A high-speed, reflective SPDT CMOS switch (Analog Devices ADG919) modulates the connection at the antenna's port to implement DAM, where one port is connected to the RF source, the other port provides the connection to the auxiliary DC source or an open circuit, depending on the case being tested. The rated switch transition time is 6.1~ns, which is considerably shorter than the 37~ns carrier period to ensure that the switch transition can be neglected. Here we test DAM with an open circuit (OC-DAM mode) or an auxiliary DC signal (DC-DAM mode) connected to the antenna during the OFF state and compare the two DAM modes with a conventional LTI transmitter (LTI mode). The switch has approximately $5~\Omega$ series resistance in the ON state, and it is left in place for all testing modes to facilitate comparison with the same bandwidth and efficiency. The total Q-factor $\eta Q_\T{rad}$ of the antenna and matching network combination is estimated to be approximately $10$. An arbitrary waveform generator (AWG, Tektronix AWG70002A) produces both the RF carrier, baseband switch control signals, and the auxiliary DC signal. 
 
 In LTI mode, the switch is held in the closed state during the entire transmission of a sequence produced by the AWG. When running in OC or DC-DAM modes, the modulated RF sequence is again produced by the AWG while the dynamic switch control as described in \cite{SchabHuangAdams2019_DAMOOK,SchabHuangAdams2020_DAMPSK} is used to toggle the switch to an open circuit or the auxiliary DC signal. Note that in this configuration, all modes radiate with identical continuous wave radiation efficiency.

\subsection{Experimental Results}
\label{sec:experimental_results}

\begin{figure}
		\centering
		\includegraphics[trim = 0.2in 0in 0.5in 0.25in,clip,width = 3.3in]{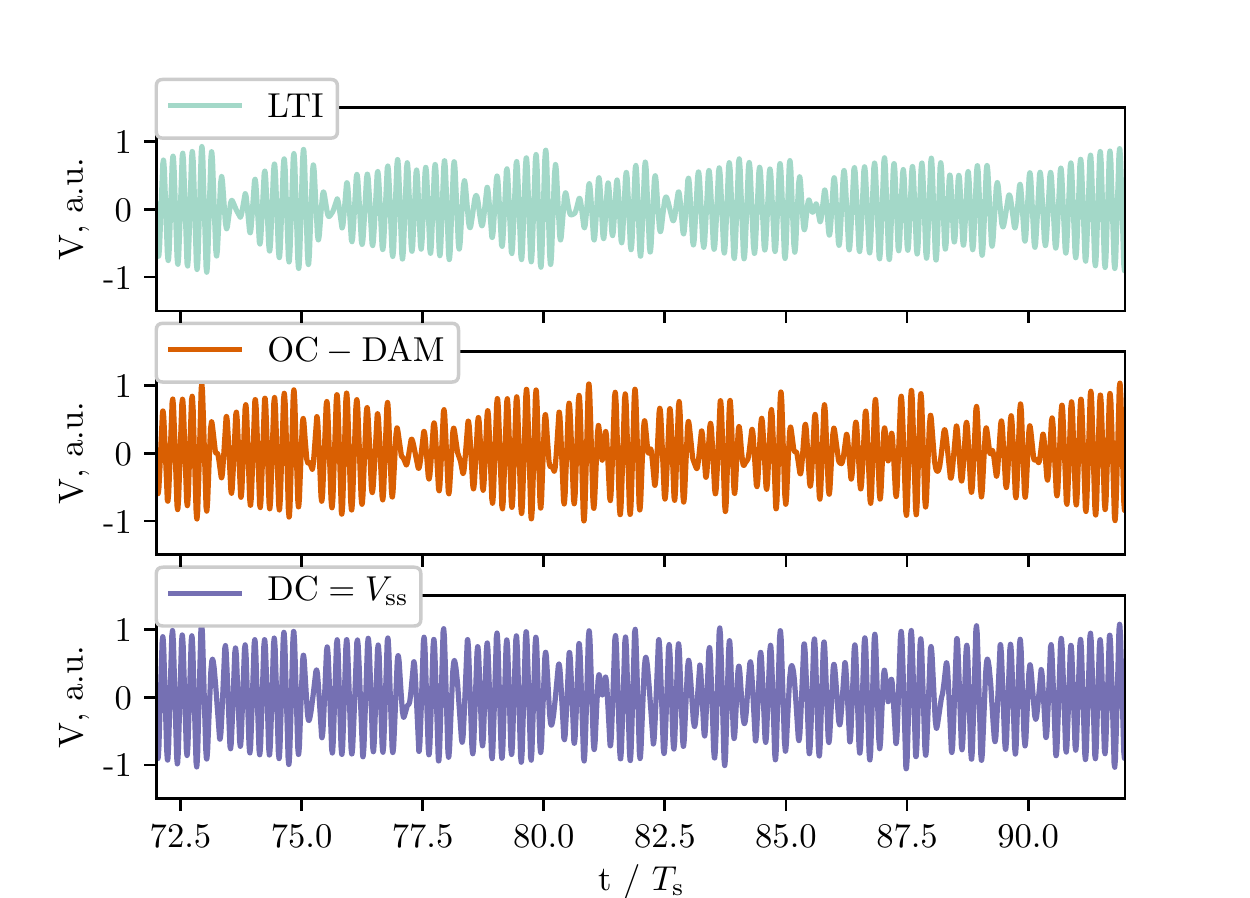}
		\caption{A portion of the received time domain signal when transmitting a QPSK PRBS with $N = 5$ in LTI, OC-DAM, and DC-DAM modes.  Time axis is scaled by symbol length $T_\mathrm{s}$.}
		\label{fig:qpsk-time-domain}
	\end{figure}

A wideband QPSK sequence is transmitted through the electrically small antenna operarting in LTI, OC-DAM, and DC-DAM modes. In DC-DAM mode, a DC voltage at the steady state amplitude is applied. A 256-bit pseudorandom bit sequence (PRBS) at two different symbol rates, 5.446 MHz and 9.077 MHz ($N=5$ and $3$, respectively) are used. This sequence is repeated 100 times, time-aligned, and averaged to create a very high signal-to-noise ratio measurement.   Fig.~\ref{fig:qpsk-time-domain}
shows a portion of the received RF signal after removing the unintentional multipath and receiver effects by a channel sounding chirp\footnote{See \cite{SchabHuangAdams2019_DAMOOK} for the details of this process.  Note that, unlike in \cite{SchabHuangAdams2019_DAMOOK}, here the receive antenna response is also inverted and removed in the reported results.} for the three different modes. We observe that the envelope of the LTI signal is much lower than either DAM case, and that some symbol transitions in OC-DAM mode exhibit longer transition time, thus producing a more distorted version of the intended PSK signal than the DC-DAM case. 

\begin{figure}
		\centering
		\includegraphics[trim = 0.22in 0.2in 0.5in 0.25in,clip,width = 3.3in]{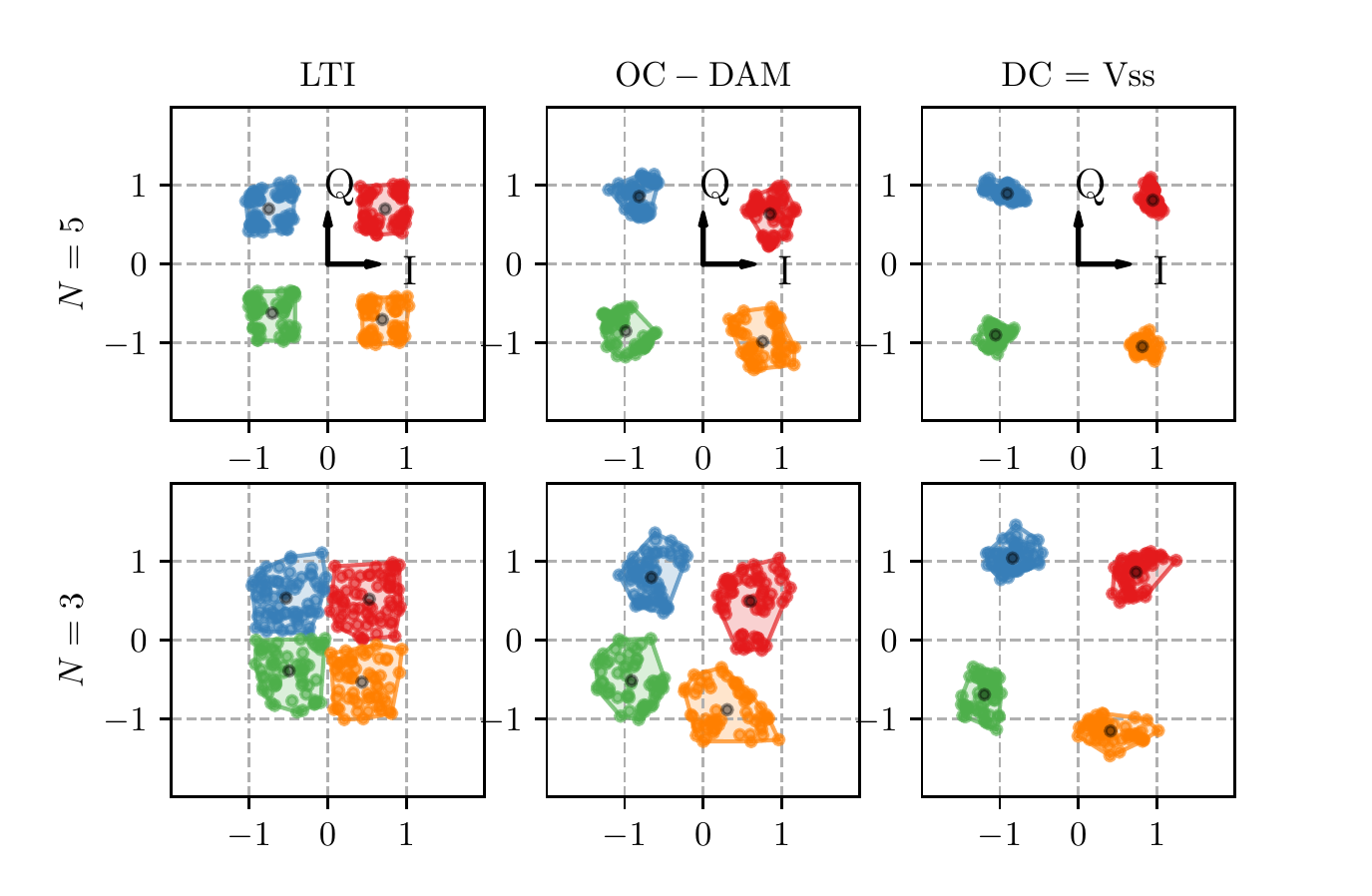}
		\caption{Measured QPSK constellations at two data rates. The four symbol states are plotted in different colors, while the mean value of each cluster is shown in black.}
		\label{fig:qpsk-const}
	\end{figure}

To evaluate signal quality more quantitatively, the measured RF data is downconverted to baseband. The measured, demodulated baseband signals are plotted on a constellation diagram in Fig.~\ref{fig:qpsk-const}
with sampling points chosen at the greatest average distance between the four symbols. Each symbol is coded with a different color and the mean value of each cluster is shown in black. As the data rate increases, the symbol clusters produced by the LTI system spread significantly due to the narrowband nature of the transmit antenna, causing symbols to overlap at the highest data rate ($N=3$). The OC-DAM case with synchronized switching spreads the clusters apart, but a noticeable spread is observed within each symbol cluster. When including the auxiliary DC source, we observe a significant reduction in cluster size and thus lower intersymbol interference. We also measure the error vector magnitude (EVM) relative to the mean value of each cluster in Fig.~\ref{fig:qpsk-const}.  


As seen in Table~\ref{table:evm}, the EVM of the conventional LTI antenna is high, particularly at the higher data rate.  The OC-DAM mode reduces the EVM by 4-5 dB, but when the proper DC signal ($V_\mathrm{DC}=V_\mathrm{ss}$) is applied in the auxiliary circuit, the EVM drops by an additional 5-9 dB.

\begin{table}
	\centering
	\caption{Error vector magnitude (dB) averaged over each symbol in the constellations of Figs.~\ref{fig:qpsk-const} and \ref{fig:dc-dam-const-diff-dc}.}
	\scalebox{1}{\begin{tabular}{ccc|ccc}
	& & & \multicolumn{3}{c}{DC-DAM ($V_\T{DC}/V_\T{ss}$)}\\
			$N$ & LTI & OC-DAM & $0.6$ & $1.0$ & $1.7$\\\hline
			5 & $-10.8$ & $-14.6$ & $-17.3$ & $-22.1$ & $-23.0$\\
			3 &$-2.3$ & $-7.0$ & $-11.3$ &$-16.2$ &$-19.5$ \\\hline
	\end{tabular}}
	\label{table:evm}
\end{table}

\begin{figure}
	\centering
	\includegraphics[trim = 0.22in 0.2in 0.5in 0.25in,clip,width = 3.3in]{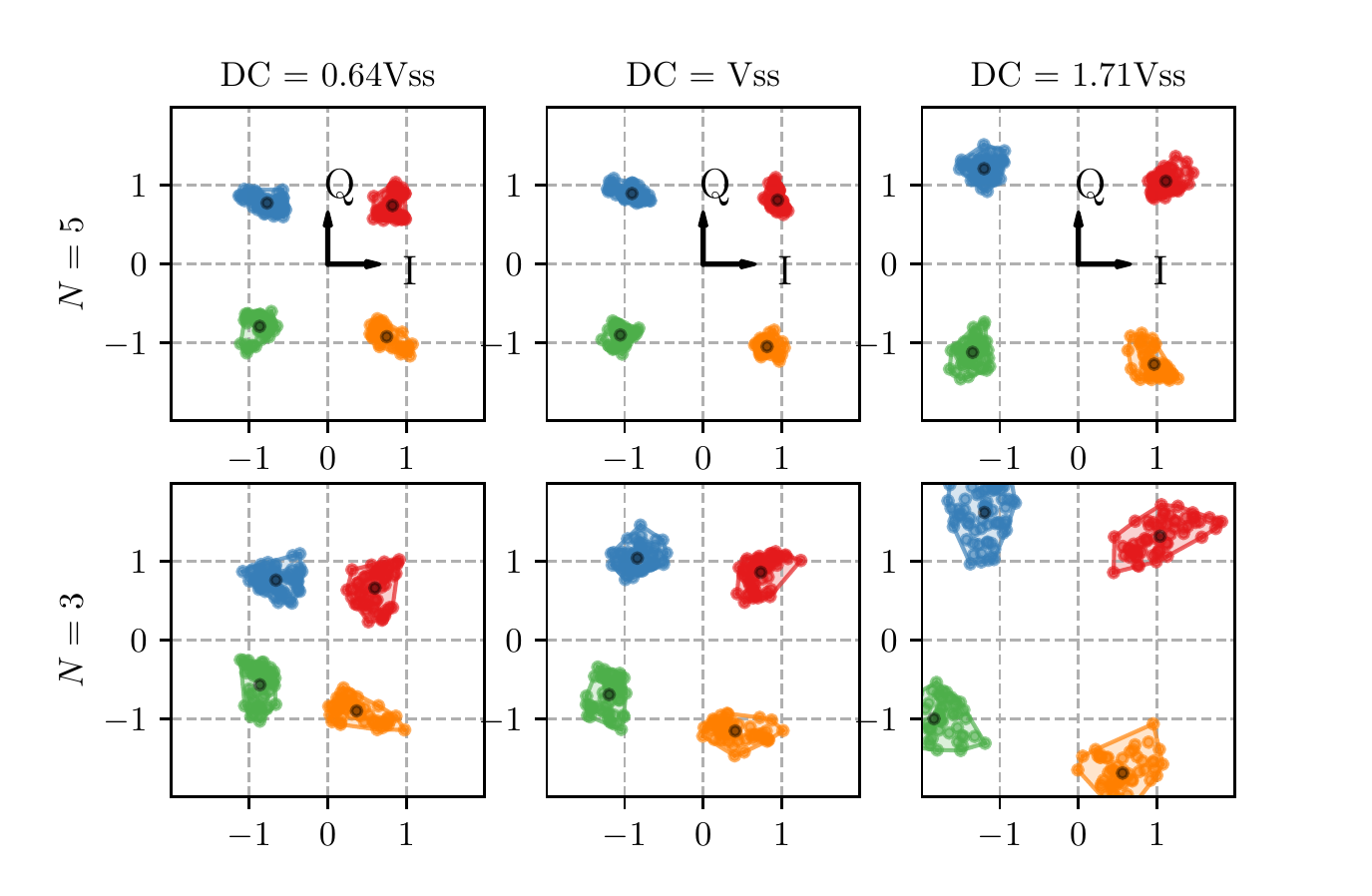}	\\
	\caption{Measured QPSK constellations when different auxiliary DC signals are applied, while the mean value of each cluster is shown in black.}
	\label{fig:dc-dam-const-diff-dc}	
\end{figure}

Additionally, we examine the DC-assisted DAM system when different DC voltages are applied to the auxiliary circuit. As is shown in the constellation diagrams in Fig.\ \ref{fig:dc-dam-const-diff-dc},
if the DC signal is decreased below the steady state value, the four symbol clusters move together with each other, reducing the noise margin. When the DC signal is increased above the steady state value, the clusters move apart as the DC source begins injecting measurable energy into the system that is upconverted and radiated. 

Table~\ref{table:evm} shows that the EVM decreases dramatically as the DC signal is increased toward the steady state value but saturates beyond that. 
\begin{table}
	\centering
	\caption{Standard Deviation (SD) of the symbols in the constellations of Figs.~\ref{fig:qpsk-const} and \ref{fig:dc-dam-const-diff-dc}.}
	\scalebox{1}{\begin{tabular}{ccc|ccc}
	& & & \multicolumn{3}{c}{DC-DAM ($V_\T{DC}/V_\T{ss}$)}\\
			$N$ & LTI & OC-DAM & $0.6$ & $1.0$ & $1.7$\\\hline
			 5 & $0.28$ & $0.26$ & $0.18$ & $0.13$ & $0.19$\\
			3 &$0.37$ & $0.38$ & $0.26$ &$0.24$ &$0.37$ \\\hline
	\end{tabular}}
	\label{table:sd}
\end{table} Table~\ref{table:sd} shows the standard deviation (SD) of the symbols from the mean of each cluster.  We observe that the standard deviation decreases when the auxiliary DC signal is increased from zero until it reaches the lowest SD when the DC voltage is equal to the steady state value. However, when the auxiliary voltage is above the steady state value, the clusters exhibit higher standard deviation, becoming more skewed and spread out over a wider range. This suggests that the DC-assisted DAM has the most compact symbol clusters when the steady state value of DC signal is applied. Overall, these results indicate significantly improved signal quality for signals transmitted with DC-assisted DAM relative to conventional systems and irrespective of further signal processing that could be applied at the receiver.




\subsection{Equivalent LTI antenna parameters}

\newcommand{\parQ}{\chi}
\newcommand{\parE}{\xi}

Measurements indicate that the proposed DC-assisted DAM method significantly improves transmitted signal quality and EVM as compared to a conventional LTI transmitter.  However, it is of general interest to quantify this improvement in terms of common LTI system parameters, \textit{e.g.}, as a bandwidth or efficiency improvement factor.  To examine DAM's effect in terms of equivalent LTI antenna parameters, we develop an analytical model of an efficiency- and bandwidth-scalable LTI antenna. Suppose that we are somehow able to independently adjust the radiation Q-factor\footnote{Many methods are available for the calculation of Q-factor, see \cite{Schab2018energy} and references therein.  Here we use an extremely simple model of radiation Q-factor proportional to the frequency derivative of an antenna's input impedance normalized to its radiation resistance assuming an overall stationary input resistance~\cite{Yaghjian2005}.} $Q_\T{rad}$ and radiation efficiency $\eta$ of an LTI antenna through the scaling parameters $\parQ$ and $\parE$, i.e.,
\begin{equation}
    \eta' = \parE \eta, \quad\quad Q_\T{rad}' = \parQ Q_\T{rad},
    \label{eq:alphabeta}
\end{equation}
Specifically, we configure the above parameterization such that the total antenna input resistance $R_\T{a} = R_\T{r}+R_\T{\Omega}$ is kept fixed, that is,
\begin{equation}
    Z_\T{a}' = R'_\T{rad} + R'_\Omega +\T{j}X'_\T{a},
\end{equation}
with
\begin{equation}
    R'_\T{rad} = \parE\eta R_\T{a},\quad R'_\T{\Omega} = (1-\parE\eta)R_\T{a},\quad X'_\T{a} = \parE\parQ X_\T{a}.
    \label{eq:alphabeta2}
\end{equation}
Using the unprimed antenna parameters $Q_\T{rad}$ and $\eta$, the transfer function $h$ between antenna input voltage and broadside radiated field is expressible as
\begin{equation}
    h = A\sqrt{\eta}(1-\varGamma),
\end{equation}
with the reflection coefficient $\varGamma$ defined as
\begin{equation}
    \varGamma = \frac{Z_\T{a}-Z_0}{Z_\T{a}+Z_0},
\end{equation}
and the multiplier $A$ being defined, to within a phase factor, by the source impedance $Z_0$, antenna input resistance $R_\T{a}$, observation distance, and radiation pattern, see \cite[\S III-B \& App. A]{schab2020distortion}.  Under the parameterization in \eqref{eq:alphabeta} and \eqref{eq:alphabeta2}, the modified transfer function with scaled radiation Q-factor and radiation efficiency is
\begin{equation}
    h'(\parE,\parQ) = h\frac{\sqrt{\parE} (1-\varGamma')}{1-\varGamma}
    \label{eq:hprime}
\end{equation}
where
\begin{equation}
    \varGamma' = \frac{R_\T{a} + \T{j}\parE\parQ X_\T{a}-  Z_0}{R_\T{a} + \T{j}\parE\parQ X_\T{a} + Z_0}.
    \label{eq:gammaprime}
\end{equation}
Note that in the altered reflection coefficient in (\ref{eq:gammaprime}), the total Q-factor of the transmitter is adjusted while maintaining the match between the source and antenna impedances at resonance, i.e., $Z_0 = R_\T{a}$ when $X_\T{a} = 0$.  Additionally, the incident power available from the source is held constant with changes in antenna impedance.

From the measured conventional antenna impedance $Z_\T{a}$, transmitted signals from a hypothetical antenna with the altered properties in \eqref{eq:alphabeta} may be obtained using \eqref{eq:hprime} and \eqref{eq:gammaprime} under varying states of the parameters $\parE$ and $\parQ$.  The same processing methods described previously can then be applied to these synthesized signals to calculate average symbol power and EVM of the parameterized model LTI antenna.

\begin{figure}
    \centering
    \includegraphics[trim = 0.25in 0.1in 0.55in 0.5in,clip,width=3.3in]{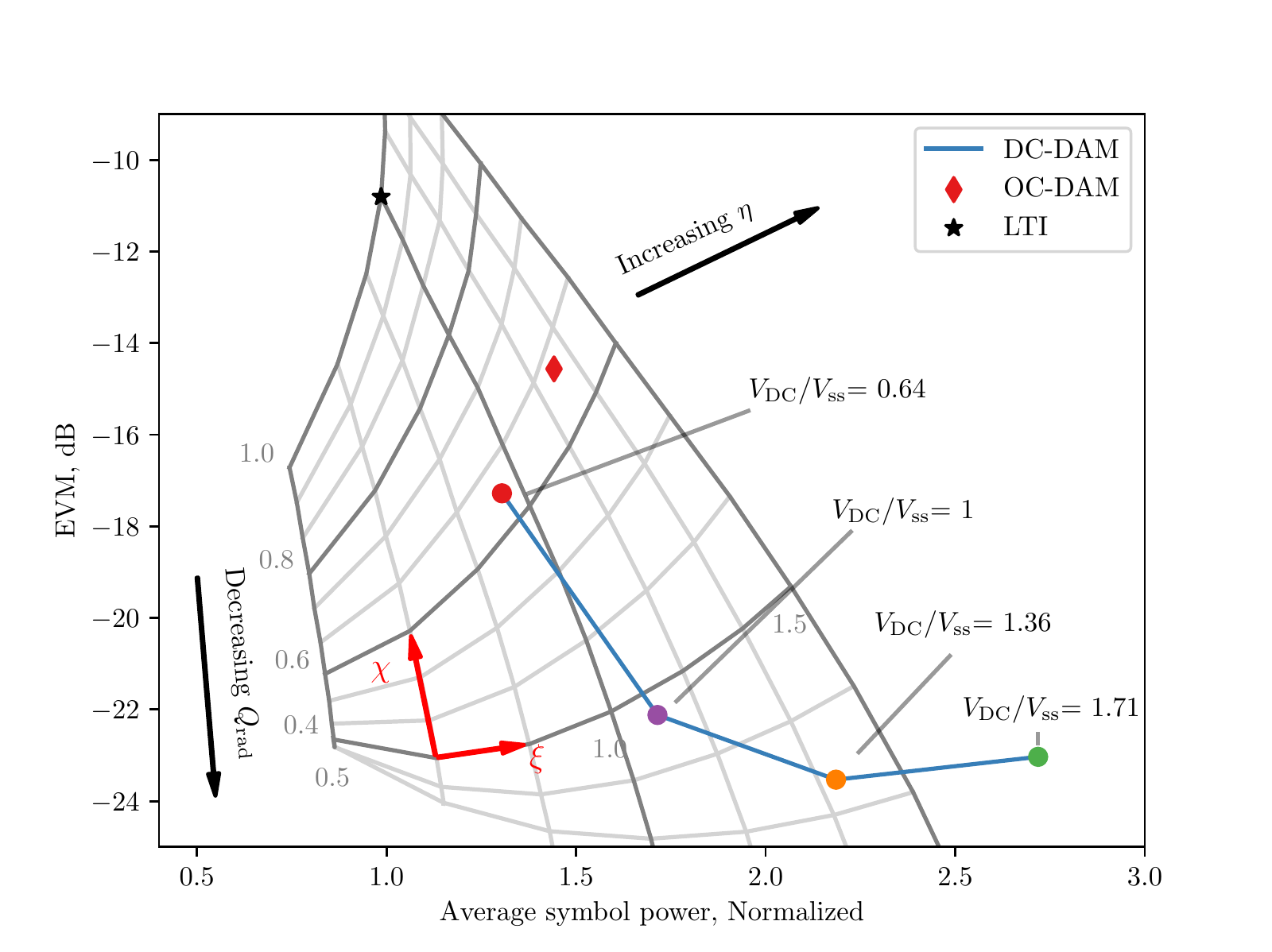}\\
    \includegraphics[trim = 0.25in 0.1in 0.55in 0.5in,clip,width=3.3in]{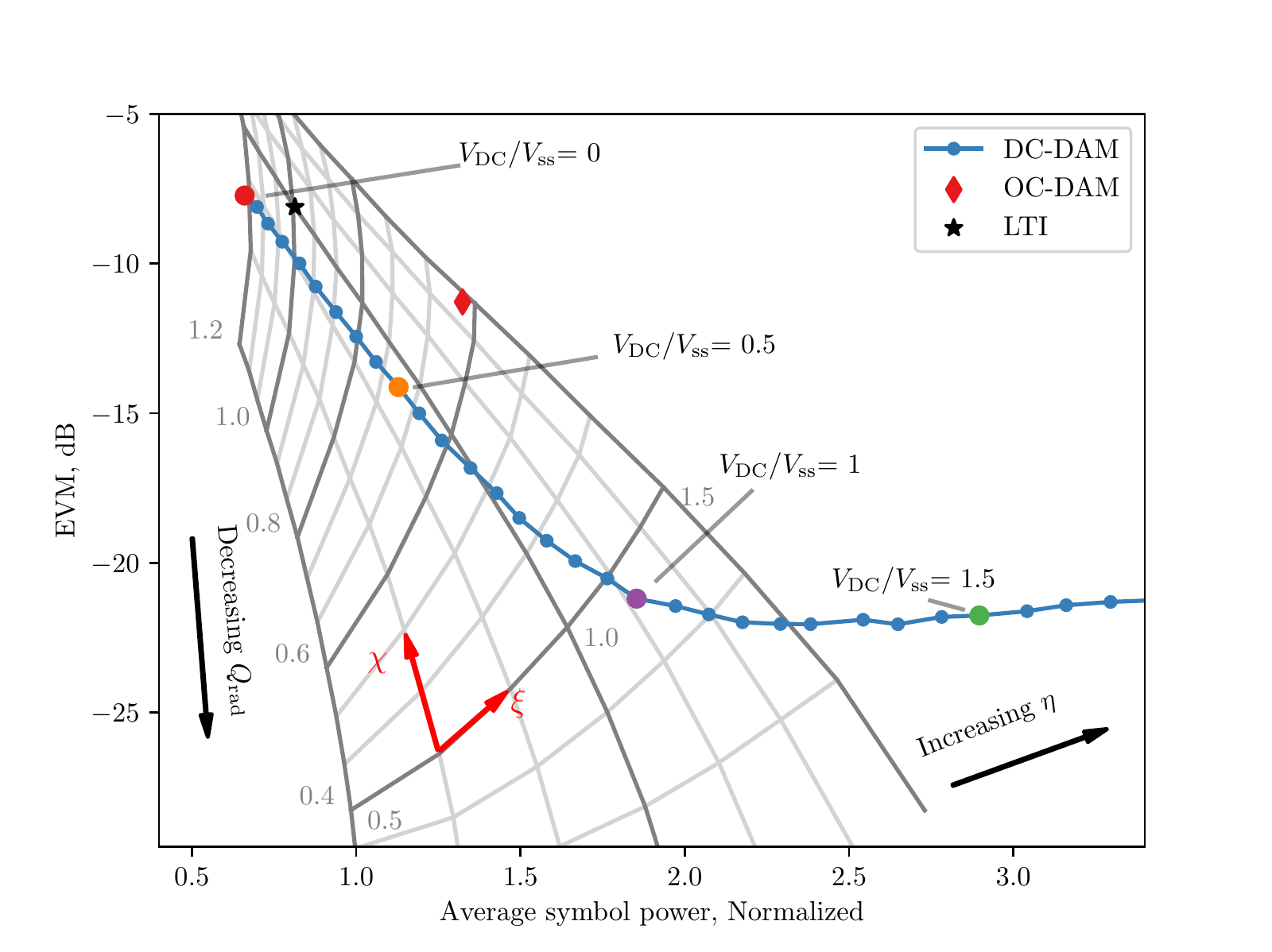}
    \caption{Comparison of DC-assisted DAM performance against conventional transmitters described by efficiency and radiation Q-factor parameters $\parE$ and $\parQ$.  Measured conventional transmitter (unaltered) corresponds to $\parE = 1$, $\parQ = 1$. The far-field measurement is shown on the top figure and the near-field lab test on the bottom.}
    \label{fig:abgrid}
\end{figure}

The characteristics of the analytical LTI antenna signals as well as the measured LTI and DAM signals are plotted in Fig.~\ref{fig:abgrid}.  Here, the $x$ and $y$ axes plot the normalized average symbol power and EVM of the received signal, respectively and thus represent signal power and signal quality.  Two important baselines are shown in Fig.~\ref{fig:abgrid}. The measured performance of the conventional LTI transmitter is shown with a black star, and the performance of the open circuit DAM transmitter is shown by a red diamond. The blue curve plots measured performance of DC-DAM transmissions with incremented auxiliary DC voltage settings with key values highlighted. The factors $\parE$ and $\parQ$ are plotted on a grid to indicate the symbol power and EVM produced by an analytical LTI transmitter with the associated scale factors. Both far-field measurements and near-field measurements with finer increments of DC voltage are shown.  The DC-DAM system follows the same trend in the far-field measurement as well as in the near-field lab test with denser data points. 

As observed in previous data, the OC-DAM case improves both signal quality and signal power over the LTI case, but the DC-DAM with $V_\mathrm{DC}=V_\mathrm{ss}$ shows significant improvements in both metrics over the LTI and OC-DAM cases.  The behavior of DC-DAM with changing DC voltage is also interesting. When the voltage source is set to zero and the antenna is shorted in the off state, both signal quality as well as signal power are reduced because the energy is discharged from the antenna rather than stored~\cite{Srivastava2017}.  As the voltage is increased from zero to the steady state value, we see that it nearly follows a line of constant efficiency but reduced radiation Q-factor.  Thus, in this regime, the DC-DAM transmitter behaves equivalently to an LTI antenna with artificially broader bandwidth and constant radiation efficiency. In fact, the effective radiation Q-factor of the DC-DAM system is $\sim 3\times$ lower than its LTI counterpart when $V_{\mathrm{DC}} = V_{\mathrm{ss}}$. Typically, such a reduction would require an equal reduction in radiation efficiency, but the effective radiation efficiency is in fact 20\% higher. When the auxiliary DC signal is raised above the steady state voltage, it brings no benefit to the EVM of the transmitted signal but does increase the radiated power because non-negligible DC power is being injected into the system. Thus, this comparison shows that signal quality improvements from DAM are equivalent to large antenna bandwidth (and to a lesser degree, radiation efficiency) improvements over an LTI transmitter.

\section{Conclusion}
The transient characteristics of a direct antenna modulation system are studied through analytical models, simulation, and experiment. Measurements confirm that parasitics of the switch and inductor induce oscillations in the OFF state of DAM system, which degrades the transition time of certain DAM modulation schemes. An alternative DAM method using an auxiliary DC signal to stabilize the stored energy on the antenna is shown to accelerate the problematic transitions in simulation and measurement. Results show that the proposed method improves the EVM of QPSK signals by 10-20 dB compared to an identical LTI transmitter. An equivalent LTI system is modeled to quantitatively assess the performance of it with DC-DAM system relative to an LTI transmitter with scalable radiation Q-factor and efficiency. Results indicate that the introduction of an auxiliary DC signal in DAM is equivalent to \textit{decreasing} the radiation quality factor of an LTI antenna by a factor of 2-3 while \textit{increasing} effective radiation efficiency. Thus, the addition of an appropriate auxiliary DC voltage to a DAM-PSK transmitter shows important benefits in reducing the effects of circuit non-idealities.


\appendix

\label{sec:append}

\subsection{Analytical expressions for DAM OFF-state behavior}
\label{subsec:derive_dam}

To predict the transient behavior of DAM circuit, we derive equations for the antenna terminal voltage $V_{\mathrm{a}}$ when the circuit is switched from the ON to the OFF state.

We first assume that the DAM circuit is switched from the ON state in Fig.~\ref{fig:dam-off}(a) to the OFF state in Fig.~\ref{fig:dam-off}(b) at the instant ($t=0$) when the antenna terminal voltage $v_{\mathrm{a}}(t)$ reaches its maximum and that the terminal current at that moment is negligible. Under these conditions, we assume that the capacitors $C$, $C_{\mathrm{s}}$, and $C_{\mathrm{L2}}$ hold an initial voltage equal to their peak steady state voltage during oscillation, while all other initial voltages and currents are small enough to be ignored. 

From the Laplace domain equivalent circuit in Fig. \ref{fig:dam-off}(c), we can write the currents generated by each source at terminal `$\T{a}$' as
\begin{equation}
V_{\mathrm{a,C}}(s)/Z_{\mathrm{a}}(s)\ =\ \displaystyle\frac{v_{\mathrm{C}}(0)/s}{Z_{\mathrm{ant}}(s)}, 
\label{eq:ini1}
\end{equation}
\begin{equation}
V_{\mathrm{a,Cs}}(s)/Z_{\mathrm{a}}(s)\ =\ \displaystyle\frac{v_{\mathrm{Cs}}(0)/s}{Z_{\mathrm{Cs}}(s)},
\label{eq:ini2}
\end{equation}
\begin{equation}
V_{\mathrm{a,CL2}}(s)/Z_{\mathrm{a}}(s)\ =\ \displaystyle\frac{v_{\mathrm{C_{L2}}}(0)/s}{Z_{\mathrm{CL2}}(s)},
\label{eq:ini3}
\end{equation}
where $Z_{\mathrm{ant}}(s)$, $Z_{\mathrm{Cs}}(s)$, and $Z_{\mathrm{C_{L2}}}(s)$ are the source impedances of the three initial voltage sources, and $Z_{\mathrm{a}}(s)$ is the total impedance from the antenna terminal `$\T{a}$' to ground. 



Expressing $V_{\mathrm{a,C}}(s)$, $V_{\mathrm{a,Cs}}(s)$, and $V_{\mathrm{a,CL2}}(s)$ as partial fraction expansions, superimposing the three sources, and converting the resulting expression to the time domain, we find the transient antenna terminal voltage

\begin{multline}
v_{\mathrm{a}}(t)\ =\ v_{\mathrm{a,C}}(t)+v_{\mathrm{a,Cs}}(t)+v_{\mathrm{a,CL2}}(t)\\ 
=\ A_{0} \mathrm{e}^{-\alpha_{0}t} - A_{1} \mathrm{e}^{-\alpha_{1}t}\sin(\omega_1t) + B_{1} \mathrm{e}^{-\alpha_{1}t}\cos(\omega_1t) \\
+ A_{2} \mathrm{e}^{-\alpha_{2}t}\sin(\omega_2t) + B_{2} \mathrm{e}^{-\alpha_{2}t}\cos(\omega_2t) ,
\label{eq:va}
\end{multline}
where the $A_{\T{n}}$ and $B_{\T{n}}$ coefficients, the $\omega_{\T{n}}$ frequencies, and the $\alpha_{\T{n}}$ decay constants are calculated numerically from the circuit components. Note that $v_{\T{a}}(t=0) = V_{\mathrm{ss}} = A_{0} + B_{1} + B_{2}$ is the steady state amplitude of the antenna terminal voltage in the ON state.



From the computed coefficients of \eqref{eq:va}, given in Table~\ref{table:coefficients}, we observe that $A_{0}$ and $B_1$ are much larger than $A_1$, $A_2$, and $B_2$ and that $\alpha_0$ is much less than $\alpha_1$ and $\alpha_2$. Thus, keeping only the dominant terms, we simplify \eqref{eq:va} as
\begin{equation}
\displaystyle v_{\mathrm{a}}(t)\ \approx\ A_{0} + B_{1} \mathrm{e}^{-\alpha_{1}t}\cos(\omega_1t),
\label{eq:va_approx}
\end{equation}
which describes a single decaying frequency tone superimposed over a DC value during the OFF state.  Substituting $V_{\mathrm{ss}}\ \approx\ A_{0} + B_{1}$ and defining $V_{\mathrm{osc}}\ =\ B_{1}$, we arrive at \eqref{eq:va_nonideal} from \eqref{eq:va_approx}.

Following the same process, the radiated field $v_{\mathrm{rad}}(t)$ of the DAM circuit in the time domain can be written as

\begin{multline}
v_{\mathrm{rad}}(t)\ =\ v_{\mathrm{rad,C}}(t)+v_{\mathrm{rad,Cs}}(t)+v_{\mathrm{rad,CL2}}(t)\\ 
=\ C_{1} \mathrm{e}^{-\alpha_{1}t}\sin(\omega_1t) - D_{1} \mathrm{e}^{-\alpha_{1}t}\cos(\omega_1t) \\
- C_{2} \mathrm{e}^{-\alpha_{2}t}\sin(\omega_2t) - D_{2} \mathrm{e}^{-\alpha_{2}t}\cos(\omega_2t). \\
\label{eq:vrad}
\end{multline}
In similar fashion, coefficients $C_{\T{n}}$ and $D_{\T{n}}$, frequencies $\omega_{\T{n}}$, and decay constants $\alpha_{\T{n}}$ can be determined numerically using the circuit elements.


Substituting the values of each circuit element into \eqref{eq:vrad} to find the coefficients in Table~\ref{table:coefficients}, we observe that the term $-D_{1} \mathrm{e}^{-\alpha_{1}t}\cos(\omega_1t)$ dominates over all other terms due to their much smaller amplitude or much higher decay rate. Thus, defining $V_{\mathrm{osc}}'\ =\ D_{1}$ into \eqref{eq:vrad}, we find the approximate relation in \eqref{eq:vrad_nonideal}.  Thus, we expect to observe a dominant decaying tone at $\omega_1$ radiating from the antenna during the OFF state. 
\begin{table}
	\centering
	\caption{Coefficients in \eqref{eq:va} and \eqref{eq:vrad} when the component values in Fig.~\ref{fig:dam-off} are used.}
	\scalebox{1}{\begin{tabular}{c|cc|cc|cc}
		$\T{n}$ & $A_{\T{n}}$ &  $B_{\T{n}}$ & $C_{\T{n}}$ & $D_{\T{n}}$ &$\omega_{\T{n}} [\T{s^{-1}}]$ &$\alpha_{\T{n}} [\T{s^{-1}}]$ \\\hline
		$0$ & $2.82$ & - & - & - & - & $2.29\times10^{3}$\\
		$1$ & $0.09$ & $0.62$ & $0.12$ & $1.09$ & $3.18\times10^{8}$ & $3.57\times10^{6}$ \\
		$2$ & $0.18$ & $0.06$ & $0.97$ & $0.06$ & $7.05\times10^{8}$ & $6.8\times10^{7}$ \\\hline
	\end{tabular}}
	\label{table:coefficients}
\end{table}

\bibliographystyle{IEEEtran}
\bibliography{dam}

\begin{thebibliography}{10}
\providecommand{\url}[1]{#1}
\csname url@samestyle\endcsname
\providecommand{\newblock}{\relax}
\providecommand{\bibinfo}[2]{#2}
\providecommand{\BIBentrySTDinterwordspacing}{\spaceskip=0pt\relax}
\providecommand{\BIBentryALTinterwordstretchfactor}{4}
\providecommand{\BIBentryALTinterwordspacing}{\spaceskip=\fontdimen2\font plus
\BIBentryALTinterwordstretchfactor\fontdimen3\font minus
  \fontdimen4\font\relax}
\providecommand{\BIBforeignlanguage}[2]{{%
\expandafter\ifx\csname l@#1\endcsname\relax
\typeout{** WARNING: IEEEtran.bst: No hyphenation pattern has been}%
\typeout{** loaded for the language `#1'. Using the pattern for}%
\typeout{** the default language instead.}%
\else
\language=\csname l@#1\endcsname
\fi
#2}}
\providecommand{\BIBdecl}{\relax}
\BIBdecl

\bibitem{Chu1948}
L.~J. {Chu}, ``{Physical Limitations of Omni-Directional Antennas},'' \emph{J.
  Appl. Phys.}, vol.~19, pp. 1163--1175, Dec. 1948.

\bibitem{manteghi2019fundamental}
M.~Manteghi, ``Fundamental limits, bandwidth, and information rate of
  electrically small antennas: Increasing the throughput of an antenna without
  violating the thermodynamic q-factor,'' \emph{IEEE Antennas Propag. Mag.},
  vol.~61, no.~3, pp. 14--26, 2019.

\bibitem{SchabHuangAdams2019_DAMOOK}
K.~R. Schab, D.~Huang, and J.~J. Adams, ``Pulse characteristics of a direct
  antenna modulation transmitter,'' \emph{IEEE Access}, vol.~7, pp.
  30\,213--30\,219, 2019.

\bibitem{SchabHuangAdams2020_DAMPSK}
------, ``An energy-synchronous direct antenna modulation method for phase
  shift keying,'' \emph{IEEE Open J. Antennas Propag.}, vol.~1, pp. 41--46,
  2020.

\bibitem{Galejs1963}
J.~Galejs, ``Switching of reactive elements in high-${Q}$ antennas,''
  \emph{IEEE Trans. Commun. Syst.}, vol.~11, no.~2, pp. 254--255, Jun. 1963.

\bibitem{Wolff1957}
H.~Wolff, ``High-speed frequency-shift keying of {LF and VLF} radio circuits,''
  \emph{IEEE Trans. Commun. Syst.}, vol.~5, no.~3, pp. 29--42, Dec. 1957.

\bibitem{Johannessen1963}
P.~Johannessen, ``Automatic tuning of high-{Q} antenna for {VLF FSK}
  transmission,'' \emph{IEEE Trans. Commun. Syst.}, vol.~12, no.~1, pp.
  110--115, Mar. 1964.

\bibitem{Hartley1971}
H.~Hartley, ``Electronic broad banding of {VLF/LF} antennas for {FSK} radio
  communication,'' \emph{IEEE Trans. Commun. Technol.}, vol.~19, no.~4, pp.
  555--561, Aug. 1971.

\bibitem{Vallese1972}
L.~Vallese, ``{VLF-LF} wide shift {FSK} antenna feed networks,'' Electrophysics
  Corp Nutley NJ, Tech. Rep., 1972.

\bibitem{Gamble1973}
J.~T. Gamble, ``Wideband coherent communication at {VLF} with the experimental
  transmitting antenna modulator ({ETAM}),'' Rome Air Development Center
  Griffiss AFB NY, Tech. Rep., 1973.

\bibitem{Xu2006}
X.~Xu, H.~Jing, and Y.~Wang, ``High speed pulse radiation from switched
  electrically small antennas,'' in \emph{IEEE Antennas Propag. Symp.}, 2006,
  pp. 167--170.

\bibitem{Xu2007}
X.~Xu and Y.~E. Wang, ``Wideband pulse transmission from switched electrically
  small antennas,'' in \emph{IEEE Radio Wireless Symp.}, 2007, pp. 483--486.

\bibitem{Xu2010}
X.~J. Xu and Y.~E. Wang, ``A direct antenna modulation (dam) transmitter with a
  switched electrically small antenna,'' in \emph{Int. Workshop Antenna
  Technol.}, 2010, pp. 1--4.

\bibitem{Zhu2014}
R.~Zhu and Y.~Ethan~Wang, ``A modified qpsk modulation technique for direct
  antenna modulation (dam) systems,'' in \emph{IEEE Antennas Propag. Symp.},
  2014, pp. 1592--1593.

\bibitem{Manteghi2016}
M.~Manteghi, ``A wideband electrically small transient-state antenna,''
  \emph{IEEE Trans. Antennas Propag.}, vol.~64, no.~4, pp. 1201--1208, 2016.

\bibitem{HuangAPS2019}
D.~Huang, J.~J. Adams, and K.~Schab, ``Ringing effects due to non-ideal
  components in direct antenna modulation transmitters,'' in \emph{IEEE
  Antennas Propag. Symp.}, 2019, pp. 1373--1374.

\bibitem{HuangAPS2020}
------, ``Investigation of ringing effects on phase shift keyed direct antenna
  modulation transmitters,'' in \emph{IEEE Antennas Propag. Symp.}, 2020, pp.
  1817--1818.

\bibitem{Salehi2014}
M.~Salehi and M.~Manteghi, ``Self-contained compact transmitter for high-rate
  data transmission,'' \emph{Electron. Lett.}, vol.~50, no.~4, pp. 316--318,
  2014.

\bibitem{WangTMTT2007}
X.~Wang, L.~P.~B. Katehi, and D.~Peroulis, ``Time-varying matching networks for
  signal-centric systems,'' \emph{IEEE Trans. Microw. Theory Techn.}, vol.~55,
  no.~12, pp. 2599--2613, 2007.

\bibitem{Schab2018energy}
K.~Schab, L.~Jelinek, M.~Capek, C.~Ehrenborg, D.~Tayli, G.~A.~E. Vandenbosch,
  and M.~Gustafsson, ``Energy stored by radiating systems,'' \emph{IEEE
  Access}, vol.~6, pp. 10\,553--10\,568, 2018.

\bibitem{Yaghjian2005}
A.~Yaghjian and S.~Best, ``Impedance, bandwidth, and q of antennas,''
  \emph{IEEE Trans. Antennas Propag.}, vol.~53, no.~4, pp. 1298--1324, 2005.

\bibitem{schab2020distortion}
K.~Schab, A.~Singh, and N.~Bohannon, ``Distortion analysis for the assessment
  of lti and non-lti transmitters,'' \emph{IEEE Trans. Antennas Propag.},
  vol.~68, no.~7, pp. 5209--5217, 2020.

\bibitem{Srivastava2017}
S.~Srivastava and J.~J. Adams, ``Analysis of a direct antenna modulation
  transmitter for wideband ook with a narrowband antenna,'' \emph{IEEE Trans.
  Antennas Propag.}, vol.~65, no.~10, pp. 4971--4979, 2017.

\end{thebibliography}

\end{document}